\documentclass{optica-article}

\journal{opticajournal} % for journals or Optica Open

\articletype{Research Article}

\usepackage{lineno}
\begin{document}

\title{Simultaneous Misalignment and Mode Mismatch Sensing in Optical Cavities Using Intensity-Only Measurements}

\author{Liu~Tao,\authormark{1,*} Eleonora~Capocasa,\authormark{1} Yuhang~Zhao,\authormark{2} Jacques~Ding,\authormark{1} Isander~Ahrend,\authormark{1} and Matteo~Barsuglia\authormark{1}}

\address{\authormark{1}Universit\'e Paris Cit\'e, CNRS, Astroparticule et Cosmologie, F-75013 Paris, France\\
\authormark{2}Institute for Gravitational Wave Astronomy, Henan Academy of Sciences, Zhengzhou, 450046, China}

\email{\authormark{*}liu.tao@apc.in2p3.fr} %% email address is required; see note below about the corresponding author designation

% use {asbstract*} to suppress the copyright line. Copyright information will be added in production

\begin{abstract*}
Precise sensing and control of spatial mode content is essential for the performance of precision optical systems, particularly interferometric gravitational-wave detectors, where misalignment and mode mismatch can lead to significant optical losses and degraded quantum noise suppression. Conventional approaches, including heterodyne wavefront sensing and phase camera techniques, are effective but can be limited by hardware complexity and systematic uncertainties arising from restricted reference-beam overlap. This paper presents a novel two-step deep learning pipeline for robust beam diagnostics based solely on beam intensity images. In the first stage, a multi-intensity-image convolutional neural network (CNN) performs accurate mode decomposition, recovering the complex modal content of distorted beams. In the second stage, the predicted mode coefficients are fed into a downstream regression network that \textit{simultaneously} estimates all eight degrees of freedom (DoFs) associated with misalignment and mode mismatch, including beam tilt, lateral offset, and waist size and position mismatches in both transverse directions. The proposed CNN-based framework achieves a mean absolute error (MAE) of 0.0034 in the mode decomposition stage, which propagates to a total MAE of 0.0062 in the recovered beam imperfection parameters at the final stage. This corresponds to an average residual optical loss of 39~ppm per DoF (310~ppm total). This approach relies only on standard CCD imaging and is robust to random intensity noise, eliminating the need for complex interferometric hardware. The results demonstrate that the proposed deep learning pipeline enables real-time, high-accuracy wavefront sensing and mode-mismatch diagnostics, providing a scalable and hardware-efficient tool for improving the stability and sensitivity of precision optical systems.

\end{abstract*}

%%%%%%%%%%%%%%%%%%%%%%%%%%  body  %%%%%%%%%%%%%%%%%%%%%%%%%%
\section{Introduction\label{sec-intro}}
Over the past decade, the U.S.-based LIGO, the European Virgo, and Japan's KAGRA gravitational-wave (GW) observatories have ushered in a new era of astrophysics by establishing GWs as a powerful observational probe, providing insights complementary to traditional electromagnetic and particle-based astronomy~\cite{Vitale_2021}. Continuous enhancements to these detectors have enabled a wide range of groundbreaking discoveries, resulting in the detection of GW signals from merging compact binary systems~\cite{GWTC-4}. Future upgrades to the current detectors are expected to push sensitivity to the limits set by the facility infrastructure~\cite{PostO5Report:2022, LIGOwhitepaper2025}. These upgrades will also serve as testbeds for key instrumentation technologies for next-generation GW observatories, such as the European Einstein Telescope (ET) and the U.S.-led Cosmic Explorer (CE)~\cite{ET_2010, CEHorizonStudy}. ET, envisioned as a 10-km interferometer with ten times the sensitivity of current detectors, will extend the GW detection horizon to redshifts beyond 100, corresponding to a time before the first stars formed. This unprecedented sensitivity will enable precision tests of fundamental physics, such as gravity, cosmology, and the properties of dense nuclear matter~\cite{ET_2010, CEHorizonStudy}.

Achieving the sensitivity goals of future gravitational-wave observatories hinges on reducing the quantum noise of the detector, which originates from the ground-state vacuum fluctuations entering the interferometer through the dark port and interfering with the circulating laser field~\cite{Caves:1980, Caves:1981}. Increasing the interferometer's operating power can suppress quantum shot noise at high frequencies, but this comes at the cost of increased radiation pressure noise at low frequencies. A well-established strategy to mitigate this noise across the full observation band is the injection of a frequency-dependent squeezed vacuum state, a quantum state of light engineered to reduce fluctuations in the measurement quadrature. After being successfully tested in prototypes~\cite{yuhangzhao_2020, FDS_MIT_2020}, Virgo and LIGO each implemented a 300-meter-long filter cavity to tailor the squeezing angle orientation across the bandwidth~\cite{Acernese_2023, Ganapathy_2023}. The use of frequency-dependent squeezing in LIGO in the last observation run (O4) allowed remarkably improved detector sensitivity and increased GW detection rates by up to 65\%~\cite{Ganapathy_2023}.

\begin{figure}[t]
    \centering
    \includegraphics[width=0.7\linewidth]{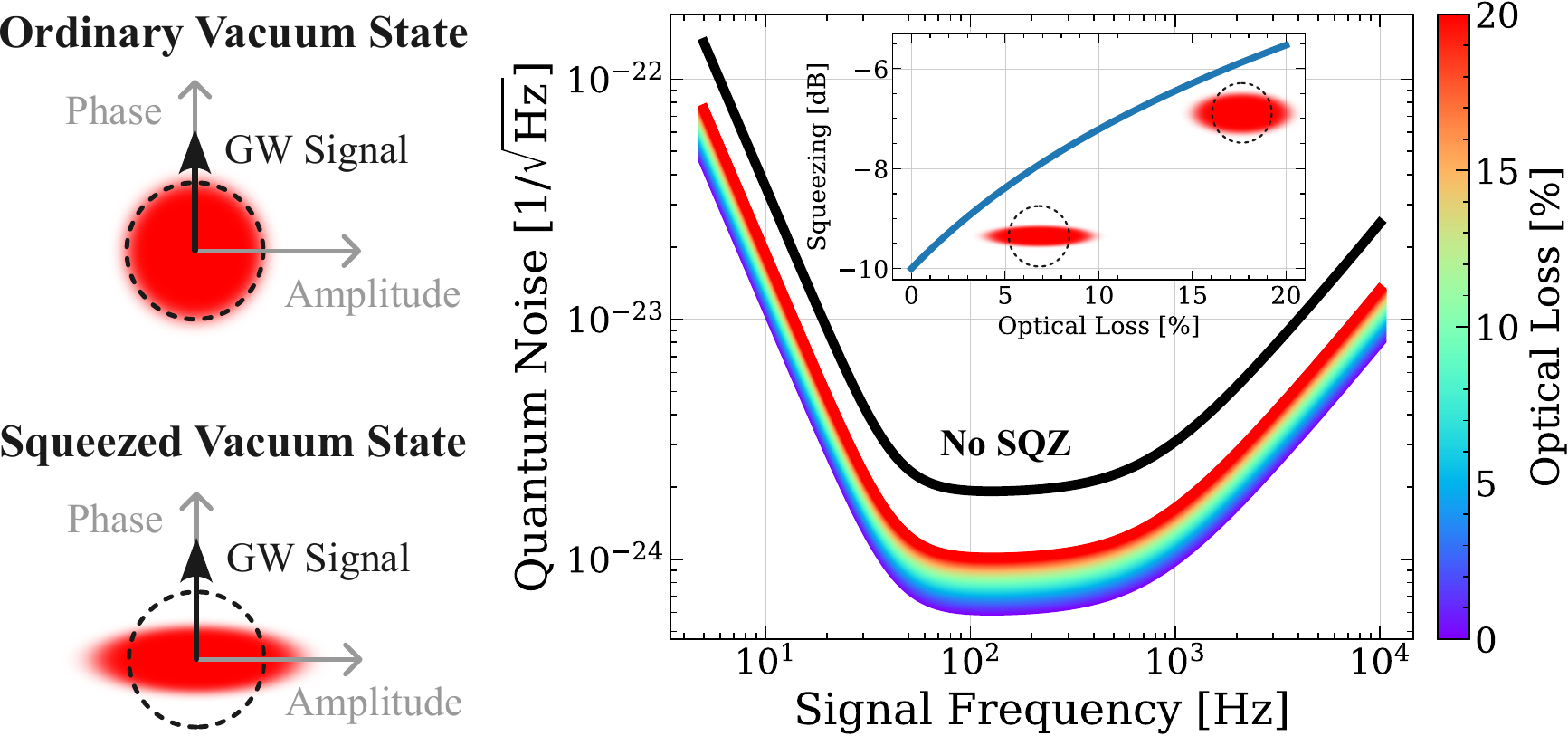}
    \caption{Impact of optical loss on observed squeezing and quantum noise spectrum. \textit{Left}: Illustration of a squeezed vacuum state, where fluctuations are suppressed in one quadrature at the expense of increased fluctuations in the orthogonal quadrature. \textit{Right}: Broadband quantum noise reduction achieved with frequency-dependent squeezing, shown as a function of increasing optical loss. This is a simplified model meant only for illustration. More realistically, depending on where they occur, most of the losses are shaped by the different cavity bandwidths in the interferometer and therefore exhibit a clear frequency dependence. The inset highlights the observed squeezing level as a function of optical loss.}
    \label{fig-QN_loss_SQZ}
\end{figure}

However, the benefits of frequency-dependent squeezing can be severely limited by various degradation mechanisms~\cite{Kwee, McCuller_2021}. The main contributor is optical losses. For instance, as illustrated in Fig.~\ref{fig-QN_loss_SQZ}, for a 10~dB injected squeezing, the effective squeezing level drops below 6~dB when a uniform 20\% optical loss is present across the frequency range. A major source of squeezing degradation is imperfect alignment and mode matching: when the input beam mode is not perfectly coupled to the cavity eigenmode, a fraction of the light is scattered into higher-order modes (HOMs) that do not carry the GW signal, directly resulting in losses. Furthermore, mismatches induce coherent scattering among HOMs, where the relative phases of the excited modes, for example, due to deviations in beam waist position or size, result in distinct interference effects. As a result, both the overall magnitude of the loss and its frequency dependence can largely change depending on the specific type of mode mismatch~\cite{Toyra_2017, Kwee, Ding25}. Because of the difficulty in accurately modeling mode mismatch and the wide variability in its impact, it has been hypothesized as the main contributor to unknown loss in current GW detectors. Reducing it therefore represents a major challenge for meeting the ambitious quantum noise reduction targets set by future upgrades and next-generation detectors.

Conventional approaches to mitigating beam misalignment and mode mismatch rely on interferometric wavefront sensing techniques such as heterodyne sensing~\cite{Morrison:94, Fabian_2019}, beam-jitter-based sensing~\cite{Fulda_17, Tao_25}, and phase cameras~\cite{Agatsuma_19, Cao_20}. While effective, these methods require complex optical and electronic hardware, including electro-optic devices, quadrant photodiodes, mode converter lenses, and radio-frequency electronics~\cite{Fabian_2019, Fulda_17, Goodwin-Jones:24, diab2025quantitativeperformanceanalysisdifferent}. Phase cameras, in particular, demand precise alignment and stability of a reference beam~\cite{Agatsuma_19}, and their limited beam size constrains the field of view, making them less suitable for large-aperture optics such as those used in gravitational-wave detectors. More recently, thermal imaging cameras have been proposed to monitor the temperature distribution and thermal state of test masses across their full aperture~\cite{tao2025FLIR}. However, such methods do not directly measure the misalignment or mode mismatch state of the interferometer's main beam and the coupled cavity system.

Determining the misalignment and mode mismatch state of a beam largely amounts to determining its higher-order modal content~\cite{Tao_21_loss}. However, mode decomposition is a challenging problem. Recently, machine learning (ML) has emerged as a powerful tool across various areas of precision optics, including laser mode generation and identification~\cite{photonics12080801, Hofer_19}, phase retrieval~\cite{Dong_2023}, automated alignment control~\cite{Qin_2025, Mukund2023}, and noise mitigation and sensitivity enhancement in gravitational-wave detectors~\cite{GWD_noise_ML, soni2025gwyolo}. For the task of mode decomposition, ML techniques using either complex-field amplitudes~\cite{Schiworski_21} or single intensity images~\cite{An_19, An_20} as input have been demonstrated. However, complex-field mode decomposition still relies on measurements of the complex amplitude obtained with a phase camera and therefore inherits the same hardware limitations. Mode decomposition method based on single intensity measurement avoids these hardware constraints but suffers from a sign ambiguity in the phase prediction~\cite{An_20}. This ambiguity is particularly critical for alignment and mode mismatch sensing, since the sign of the phase directly determines the sign of the beam imperfection and the associated error signal.

Here, we introduce a novel two-step machine learning framework for robust beam diagnostics to address imperfect beam alignment and mode matching. In the first step, we apply an intensity-image-based mode decomposition via convolutional neural networks (CNN) that utilizes multiple beam intensity images measured at different propagation planes with large Gouy phase separation. The phase ambiguity present in the single-intensity-image approach is eliminated by using multiple images captured in both the near and far fields~\cite{Cutolo_95}. In the second step, the reconstructed complex mode coefficients serve as inputs to a downstream deep learning regression model that predicts all degrees of freedom (DoFs) associated with generic misalignment and mode mismatch. This includes beam tilt and lateral offset, which describe axis misalignment, as well as waist size and waist position mismatches in both transverse directions, resulting in eight DoFs in total.

This intensity-image-based mode decomposition technique provides a cost-effective and compact solution for beam diagnostics, requiring only standard CCD cameras. Once trained, the ML models can be evaluated with minimal computational cost, enabling \textit{real-time} wavefront sensing and \textit{simultaneous} beam diagnostics of generic misalignment and mode mismatch. This represents a significant improvement over conventional interferometric sensing methods, which typically rely on linearized error signals valid only under small perturbations and require separate sensors and optical paths to address alignment and mode-matching errors independently. 

Our two-step pipeline is designed such that the mode decomposition itself serves as a standalone tool with broad applicability beyond alignment and mode mismatch diagnostics. For example, this approach could be extended to higher-order wavefront aberration correction, such as spherical aberrations~\cite{Wang:21}. We further demonstrate that the CNN-based mode decomposition model can also serve as an effective beam intensity ``denoiser'', removing speckle artifacts and realistic measurement noise in the reconstructed intensity. This work demonstrates that the proposed two-step ML pipeline can achieve a robust prediction of misalignment and mode mismatch with a final mean absolute error of 0.0062. This corresponds to approximately 310~ppm of total residual optical loss across all eight orthogonal DoFs. This capability opens a realistic pathway towards real-time diagnostics of beam perturbations in large-optics experiments, such as gravitational-wave detectors.

The paper is organized as follows: in \S\ref{sec-background}, we review the relevant background on the role of mode decomposition in diagnosing misalignment and mode mismatch. In \S\ref{sec-setup}, we describe the proposed ML framework in detail, including data preparation, model architecture, and training and testing. Section~\S\ref{sec-result} presents the results of the two-step ML pipeline, and \S\ref{sec-conclusion} concludes with an overview of the broader implications and potential extensions of this work.

\section{Background \label{sec-background}}

\subsection{Hermite-Gaussian Mode Decomposition}

\begin{figure}[b]
    \centering
    \includegraphics[width=0.5\linewidth]{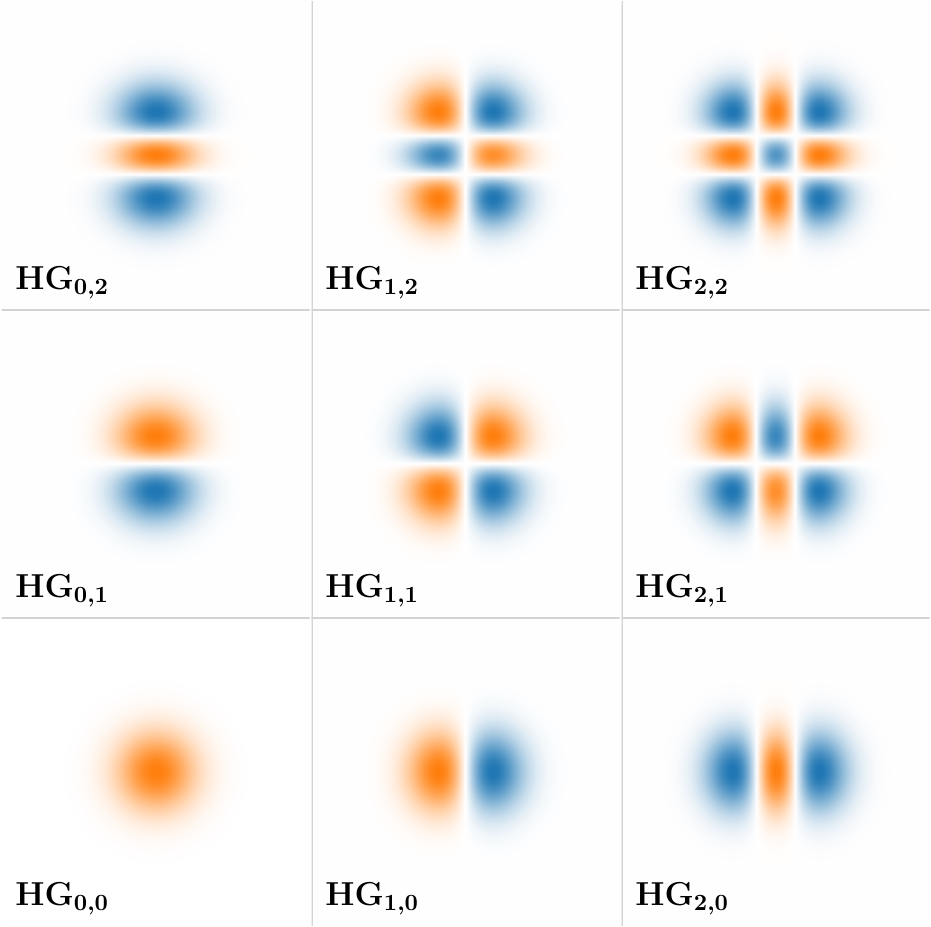}
    \caption{Amplitudes of Hermite-Gaussian modes, $\mathrm{HG}_{n,m}$, evaluated at the beam waist for mode indices $n,m \leq 2$. The higher-order modes, beyond the fundamental $\mathrm{HG}_{0,0}$, arise from beam misalignment and mode mismatch within the system.}
    \label{fig-HOMList}
\end{figure}

Hermite-Gaussian (HG) modes form a complete orthonormal basis of solutions to the paraxial Helmholtz wave equation in Cartesian coordinates~\cite{Bond2017}. They provide an accurate description of free-space Gaussian beam propagation and serve as an excellent approximation for the eigenmodes of optical resonators, particularly in the presence of astigmatism. Each HG mode is characterized by a pair of non-negative integers, $n$ and $m$, which define the mode indices and correspond to the number of nodes along the $X$ and $Y$ transverse directions, respectively. An individual mode is denoted as $\mathrm{HG}_{n,m}$. The general expression for an $\mathrm{HG}_{n,m}$ mode can be written as

\begin{equation}
\mathcal{U}_{n m}(x, y, z)=\mathcal{U}_{n}(x, z) \mathcal{U}_{m}(y, z) ,
\end{equation}
with 
\begin{equation}
\begin{aligned} 
\mathcal{U}_{n}(x, y, z)= & \left(\frac{2}{\pi}\right)^{1 / 4}\left(\frac{\exp (i(2 n+1) \Psi(z))}{2^{n} n!w(z)}\right)^{1 / 2} \\ & \times H_{n}\left(\frac{\sqrt{2} x}{w(z)}\right) \exp \left(-i \frac{k x^{2}}{2 R_{c}(z)}-\frac{x^{2}}{w^{2}(z)}\right) ,
\end{aligned}
\end{equation}
where $k$ is the wavenumber, $\lambda$ is the wavelength, $w(z)$ is the beam radius as a function of longitudinal position, and $R_{c}(z)$ is the wavefront radius of curvature. The term $\Psi(z)=\arctan \left(\frac{z-z_{0}}{z_{R}}\right)$ is the Gouy phase, where $z_{R} = \frac{\pi w_{0}^2}{\lambda}$ is the Rayleigh range, $z_{0}$ is the waist location, and $w_{0}=w(z=z_{0})$ is the beam waist size. The function $H_{n}(x)$ denotes Hermite polynomials of order $n$, containing $n$ transverse nodes. These nodal structures give rise to the characteristic multi-lobed amplitude distributions for higher-order HG modes, as illustrated in Fig.~\ref{fig-HOMList} for the case $n,m \leq 2$. The mode order of a Hermite-Gaussian beam is conventionally defined as $n+m$.

An HG mode basis is uniquely specified by its waist sizes ($w_{0,x}$ and $w_{0,y}$) and waist locations ($z_{0,x}$ and $z_{0,y}$), allowing for different beam parameters along the two transverse directions $X$ and $Y$. A convenient formalism for describing the propagation properties is through the complex beam parameter,
\begin{equation}
q=z-z_{0} + i \cdot z_{R} .
\end{equation}
which compactly encodes both the waist location and size through its real and imaginary parts.

When a laser beam in the fundamental Gaussian mode couples into an optical system, such as an optical cavity, in the presence of imperfections, the coupled field can be expressed as a superposition of higher-order HG modes in the basis of the optical system
\begin{equation}
\mathcal{U}_{0, 0}(x, y, \tilde{q})=\sum_{n=0}^{\infty} \sum_{m=0}^{\infty} c_{n,m} \cdot \mathcal{U}_{n,m}(x, y, q_{0}) ,
\end{equation}
where the eigenmode of the cavity basis that the beam is coupling to is denoted as $q_{0}$ and the mode of the incident beam itself is denoted as $\tilde{q}$. The mode coefficients $c_{n,m}$ are, in general, complex, representing both the magnitude and relative phase of the scattered HOMs.

The mode coefficients can be solved by exploiting the orthonormality of the HG mode basis
\begin{equation}
c_{n,m}=\int^{\infty}_{-\infty} \int^{\infty}_{-\infty} \mathrm{d} x \mathrm{d} y ~ \mathcal{U}_{0, 0}(x, y, \tilde{q}) \cdot \mathcal{U}_{n,m}^{*}(x, y, q_{0}) .
\label{eq-overlap}
\end{equation}

The HG mode decomposition is the process of determining these complex mode coefficients $c_{n,m}$, which quantify the contribution of each mode to the perturbed beam.

\subsection{Misalignment and Mode Mismatch}

When a laser beam in the fundamental Gaussian mode couples into an optical cavity whose eigenmodes are well described by the Hermite-Gaussian basis, imperfections in the coupling process can arise from both misalignment and mode mismatch. Misalignment occurs when the beam axis deviates from the cavity eigen-axis, either through lateral offset or an angular tilt at the beam waist. Mode mismatch, on the other hand, arises from discrepancies between the waist parameters of the incident beam and those of the cavity eigenmode, specifically, differences in the waist size and waist position. 

\begin{figure}[htbp]
    \centering
    \includegraphics[width=0.6\linewidth]{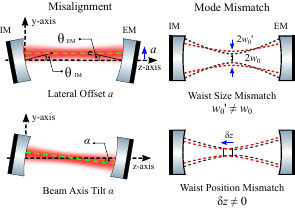}
    \caption{Illustration of beam misalignment (left) and mode mismatch (right) when coupling a laser beam into an optical cavity. Misalignment is described by two orthogonal DoFs: lateral offset and beam axis tilt. Mode mismatch is similarly described by two orthogonal DoFs: waist size mismatch and waist position mismatch.}
    \label{fig-illustration_misalignment_MM}
\end{figure}

For instance, as illustrated in Fig.~\ref{fig-illustration_misalignment_MM}, the beam axis (shown in green) can be laterally displaced by an offset $a$ from the cavity axis (shown in black) when the input and end mirrors (IM and EM) are rotated in opposite directions. Conversely, a tilt $\alpha$ arises at the cavity waist when both mirrors are rotated in the same direction. Similarly, mode mismatch can be represented by a difference in the beam waist size $w_{0}'$ relative to the cavity eigenmode waist $w_{0}$ at the cavity waist, or by a longitudinal offset $\delta z$ between the two waist positions. Thus, in total, there are eight independent imperfection parameters, accounting for both transverse directions.

In the presence of such perturbations, the coupled field no longer remains purely in the fundamental mode but instead couples into higher-order HG modes of the cavity basis. Specifically, when the imperfections are small, and the beam remains nearly matched to the cavity, misalignment primarily couples light from the fundamental mode into first-order modes, while mode mismatch mainly excites second-order modes~\cite{Tao_21_loss}. This modal scattering directly leads to measurable losses, degraded cavity buildup, and reduced interferometric contrast. Accurately quantifying and compensating these misalignment and mode mismatch parameters are therefore essential for optimal coupling and maintaining interferometric stability and sensitivity in precision optical systems such as gravitational-wave detectors.

\begin{table}[b]
\centering
\caption{Parameters describing orthogonal misalignment and mode mismatch DoFs, their leading scattered higher-order modes, and associated power loss. Misalignment is defined by tilt and offset, while mode mismatch is defined by waist position (WP) and waist size (WS) mismatch. All imperfections are assumed along the $X$ direction, showing only $\mathrm{HG}_{1,0}$ and $\mathrm{HG}_{2,0}$; equivalent results hold for the orthogonal $Y$ direction.}
\begin{tabular}{c|c|c|c}
\hline 
\hline 
Imperfections & Parameter & \hspace{0.3cm} Leading HOM \hspace{0.3cm} & Power Loss \\
\hline
Tilt & $\alpha$ & $i \frac{\alpha}{\Theta} \cdot \mathrm{HG}_{1,0}$ & $ \left(\frac{\alpha}{\Theta} \right)^{2}$ \\ \hline 
Offset & $a$ & $ \frac{a}{w_{0}} \cdot \mathrm{HG}_{1,0}$ &  $ \left(\frac{a}{w_{0}} \right)^{2}$ \\ \hline 
WP & $\delta z$ & $i \frac{\sqrt{2}}{4}\frac{\delta z}{z_{R}} \cdot \mathrm{HG}_{2,0}$ & $ \frac{1}{8} \left(\frac{\delta z}{z_{R}}\right)^{2} $ \\ \hline 
WS & $\delta w_{0}$ & $ \frac{\sqrt{2}}{2} \frac{\delta w_{0}}{w_{0}}\cdot  \mathrm{HG}_{2,0}$ &  $ \frac{1}{2} \left(\frac{\delta w_{0}}{w_{0}}\right)^2$ \\ 
\hline \hline 
\end{tabular}
\label{tab-MAMM}
\end{table}

The leading contributions of each perturbation to the scattered modes are summarized in Tab.~\ref{tab-MAMM}~\cite{Tao_21_loss}. For convenience, we define the following normalized parameters characterizing various beam imperfections:
\begin{equation}
\begin{array}{rlrl}
\epsilon_{\alpha} &= \frac{\alpha}{\Theta}, & \epsilon_{a} &= \frac{a}{w_{0}}, \\[4pt]
\epsilon_{z} &= \frac{\sqrt{2}\delta z}{4z_{R}}, & \epsilon_{w} &= \frac{\sqrt{2}\delta w_{0}}{2w_{0}},
\end{array}
\end{equation}
for the orthogonal DoFs in misalignment (top row) and mode mismatch (bottom row). $\Theta = \frac{\lambda}{\pi w_{0}}$ is the beam far-field divergence angle. In this way, the leading HOM amplitudes and the resulting power loss can be expressed in a simple form
\begin{equation}
\begin{array}{rcl}
\text{HOM amplitude:} &\quad& \epsilon \text{ or } i\epsilon, \\[6pt]
\text{Power loss:} &\quad& \epsilon^2 ,
\end{array}
\end{equation}
depending on the type of misalignment or mismatch. For instance, for lateral offset and waist size mismatch, the leading HOM amplitude is 
$\epsilon$, whereas for tilt and waist position mismatch, the leading amplitude is $i\epsilon$.

However, as the magnitude of imperfections grows beyond the linear regime, the next order terms in the expansion series become increasingly significant, causing the leading HOM amplitudes and associated power losses to deviate from the linear approximations given in Tab.~\ref{tab-MAMM}. Furthermore, the contributions from orthogonal misalignment and mode mismatch DoFs begin to mix. For example, a large tilt can generate higher-order terms corresponding to real second-order modes, which mimic the effect of waist size mismatch. Similarly, simultaneous waist position mismatch and lateral offset can produce imaginary first-order modes that resemble tilt. Consequently, when misalignment and mode mismatch occur simultaneously at large amplitudes, the scattering into first- and second-order modes becomes a highly cross-coupled, non-linear problem. Capturing this intricate mode mixing is precisely where machine learning excels.

\begin{figure}[t]
    \centering
    \includegraphics[width=0.7\linewidth]{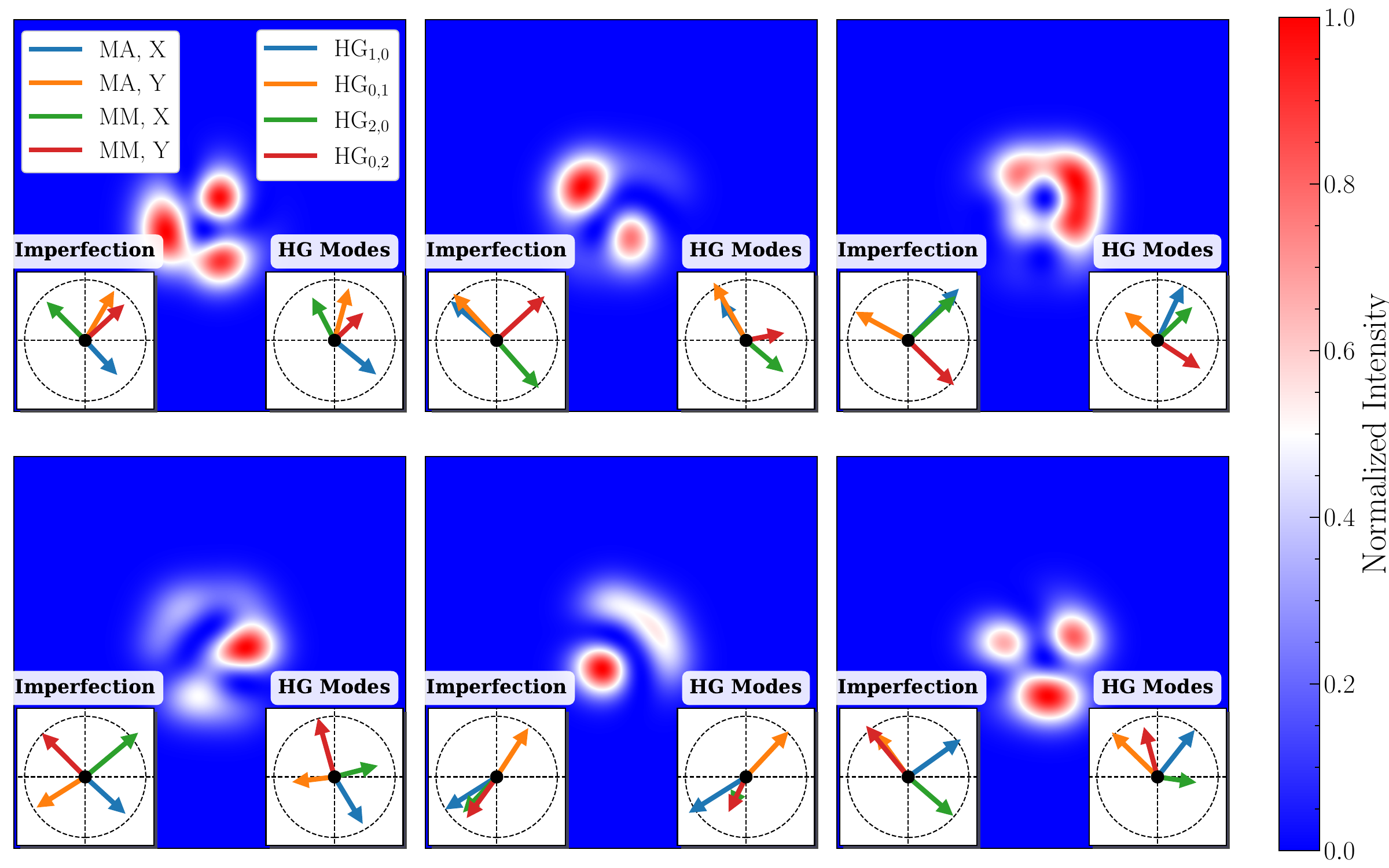}
    \caption{Example intensity images showing scattered modal content from random misalignment and mode mismatch. Six representative cases are shown. The lower-left subpanel (titled ``Imperfection'') summarizes the applied misalignment (MA) and mode mismatch (MM) along the $X$ and $Y$ directions: the horizontal axis represents lateral offset (MA) and waist size mismatch (MM), while the vertical axis represents beam tilt (MA) and waist position mismatch (MM). The lower-right subpanel (titled ``HG Modes'') shows the resulting first-order modes ($\mathrm{HG}_{1,0}$ and $\mathrm{HG}_{0,1}$) and second-order modes ($\mathrm{HG}_{2,0}$ and $\mathrm{HG}_{0,2}$), with axes corresponding to the real and imaginary parts of the complex mode amplitudes.}
    \label{fig-HOM_misalignment_MM}
\end{figure}

Fig.~\ref{fig-HOM_misalignment_MM} presents six example intensity images illustrating the total scattered modal content resulting from random misalignment and mode mismatch. For each case, two subpanels are included: one showing phasors representing the beam imperfection parameters, while the other shows phasors representing the corresponding scattered HG mode components. The lower-left subpanel, titled ``Imperfection'', indicates the type of misalignment or mode mismatch. The horizontal axis represents lateral offset $\epsilon_{a}$ (misalignment) and waist size mismatch $\epsilon_{w}$ (mode mismatch), while the vertical axis represents beam tilt $\epsilon_{\alpha}$ (misalignment) and waist position mismatch $\epsilon_{z}$ (mode mismatch). The imperfection parameters $\vec{\epsilon}$ are drawn from a uniform distribution on $[1/8,\, 1/4]$ with randomly assigned sign, ensuring that the injected imperfections are comparatively large. For reference, a circle of radius $\sqrt{2}/4$ is shown, encompassing all imperfection vectors generated by the specified distribution.

The lower-right subpanel, titled ``HG Modes'', displays the resulting first-order modes ($\mathrm{HG}_{1,0}$ and $\mathrm{HG}_{0,1}$) and second-order modes ($\mathrm{HG}_{2,0}$ and $\mathrm{HG}_{0,2}$), with the horizontal and vertical axes corresponding to the real and imaginary parts of the complex mode amplitudes. Similarly, a circle of radius $\sqrt{2}/4$ is shown to indicate the relative amplitude of the HG mode components.

For small imperfections, only the leading HOM contribution is significant, and according to Tab.~\ref{tab-MAMM}, the corresponding arrows in the left and right panels should have the same length and direction. However, as Fig.~\ref{fig-HOM_misalignment_MM} demonstrates, this is no longer true when large misalignment and mode mismatch occur simultaneously. This shows that accurately disentangling the effects of misalignment and mode mismatch for simultaneous beam diagnostics based on higher-order mode scattering is generally a complex and nontrivial task. In this work, we address this challenge by employing a machine learning model pipeline specifically designed for this purpose.

\section{Model Setup  \label{sec-setup}}

\begin{figure*}[t]
    \centering
    \includegraphics[width=0.8\linewidth]{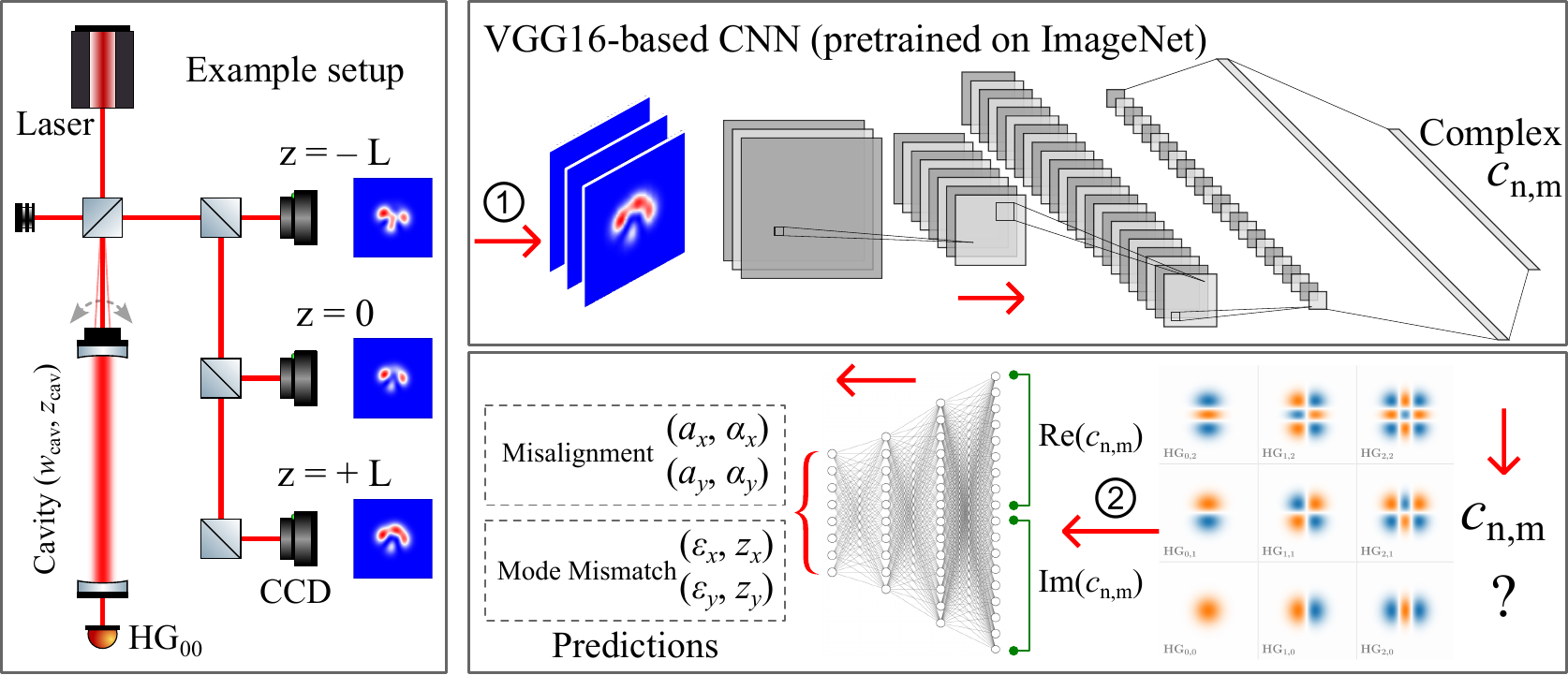}
    \caption{Schematic of the two-step machine learning pipeline for mode decomposition and beam diagnostics. In the first stage, an adapted VGG16-based convolutional neural network (CNN) takes three beam intensity images measured at different locations as input and outputs the complex mode amplitudes up to order 2. In the second stage, these mode coefficients are fed into a separate neural network that predicts all eight misalignment and mode mismatch degrees of freedom simultaneously.}
    \label{fig-concept_fig2}
\end{figure*}

To address the complex beam diagnostic problem for misalignment and mode mismatch using only intensity images, we adopt a two-step machine learning pipeline. The first step involves training a convolutional neural network that takes as input a set of three intensity images measured at different propagation distances in the reflection of an optical cavity, as illustrated in the example setup in Fig.~\ref{fig-concept_fig2}. While the implementation described here uses multiple CCD cameras at different propagation distances, the method fundamentally requires only a set of intensity measurements. In practice, this can be achieved using a single camera on a translation stage or via a compact imaging system to access multiple effective planes, reducing the overall experimental footprint.

We consider an impedance-matched, high-finesse optical cavity that is maintained on resonance for the fundamental mode. Under this condition, cavity leakage is negligible, and the transmitted field contains the remaining fundamental mode, while the reflected field contains the scattered HOMs. Once trained under these conditions, the model can be readily adapted to other cavity configurations and layouts via transfer learning, enabling rapid deployment without full retraining. For instance, in a test case with an alternative cavity configuration where a large fraction of the fundamental mode is present in reflection on top of the scattered HOMs, a single-stage transfer learning is sufficient to fully recover the original CNN accuracy, efficiently learning both the fundamental and HOM amplitudes. The CNN outputs the complex mode coefficients for HG modes with $n,m \leq 2$, whose corresponding amplitude profiles are shown in Fig.~\ref{fig-HOMList}.

Eight higher-order modes are included, corresponding to sixteen independent mode coefficients, since each coefficient is generally complex. This mode set includes two third-order modes, $\mathrm{HG}_{1,2}$ and $\mathrm{HG}_{2,1}$, and one fourth-order mode, $\mathrm{HG}_{2,2}$, which can arise in the presence of large simultaneous misalignment and mode mismatch. For example, $\mathrm{HG}_{1,2}$ can result from a combination of large misalignment in the $X$ direction and mode mismatch in the $Y$ direction. Including these modes above the second order improves the ability to decouple orthogonal beam imperfections. We refer to this first step as the ``CNN model'', as illustrated in Fig.~\ref{fig-concept_fig2}.

In the second step, we use the predicted HOM amplitude coefficients from the upstream CNN model together with the fundamental mode amplitude transmitted through the cavity to train a deep learning (DL) based regression model. This model, implemented as a densely connected multi-layer perceptron, predicts the eight orthogonal misalignment and mode mismatch parameters simultaneously. We refer to this step as the ``DL model'', completing the two-step ML pipeline for robust beam diagnostics.

\subsection{Data Preparation}

\begin{table}[b]
\centering
\caption{Key parameters assumed in generating the dataset for training, validating, and testing of the machine learning model.}
\setlength{\tabcolsep}{14pt}
\begin{tabular}{c|c}
\hline \hline
Parameter & Value \\
\hline
Waist Size & 500 $\mu m$ \\
\hline
Image Width & 4 $mm$ \\
\hline
Pixels & 128 \\
\hline
Camera Locations & z=[-0.74 $m$, 0,  0.74 $m$] \\
\hline
Imperfections & $-1/4 \leq \epsilon \leq 1/4$ \\
\hline \hline
\end{tabular}
\label{tab-data_param}
\end{table}

Training a machine learning model for a complex multi-image-based mode decomposition and beam diagnostics task requires a large and diverse dataset. We generate this dataset numerically by simulating the complex Hermite-Gaussian mode amplitudes as two-dimensional arrays. Random misalignment and mode mismatch are introduced to the fundamental mode amplitude, which is then numerically decomposed into the HG mode basis to obtain the complex mode coefficients. The corresponding intensity images of the HOM content are reconstructed at various locations.

Tab.~\ref{tab-data_param} summarizes the set of parameters assumed for the dataset preparation. We choose the waist size of both the unperturbed beam and the cavity eigenmode to be $500\ \mu\text{m}$, a value typical in precision optics experiments. The longitudinal locations for CCD imaging are set at the beam waist and $\pm z_R$ from the waist, where $z_R = 0.74\ \text{m}$ is the Rayleigh range. This arrangement produces sufficient Gouy phase evolution ($\pi/4$ per mode order) between the consecutive intensity measurements, providing adequate phase diversity to facilitate phase recovery from beam intensity measurements alone. Additionally, these measurement locations leave enough physical space to set up beam splitters and CCD cameras, as illustrated in the example experimental setup in Fig.~\ref{fig-concept_fig2}. The image width is chosen to be eight times the waist size to ensure the entire beam profile is captured at the camera positions with negligible clipping loss.

We randomly sample misalignment and mode mismatch across all eight independent DoFs, $\epsilon_{\alpha, x/y}$, $\epsilon_{a, x/y}$, $\epsilon_{z, x/y}$, and $\epsilon_{w, x/y}$, covering both transverse directions using a uniform distribution. The sampling range for each parameter is set to 
$[-1/4, 1/4]$, ensuring that the worst-case total power loss scattered from the input fundamental Gaussian mode is approximately
\begin{equation}
8 \cdot \left( \frac{1}{4} \right)^2 = 0.5,
\end{equation}
i.e., about $50\%$ of the total input power.

For each randomly generated combination of imperfections, the perturbed beam amplitude is numerically decomposed in the original HG mode basis by computing the overlap integral in Eq.~(\ref{eq-overlap}), yielding the complex mode coefficients. The common phase factor of the mode coefficients is removed, such that the phases are expressed relative to the phase of the fundamental mode, since only the relative phase between modes carries physical significance. These extracted higher-order mode coefficients are then used to reconstruct the amplitude and intensity profiles of the beam reflected from the cavity at the three distinct positions along the beam propagation direction. Each intensity profile has dimensions of $128\times128$ pixels. We generate a dataset comprising $80,000$ samples for training and validation, and a separate set of $1,000$ samples reserved exclusively for testing the model.

\begin{figure}[t]
    \centering
    \includegraphics[width=0.7\linewidth]{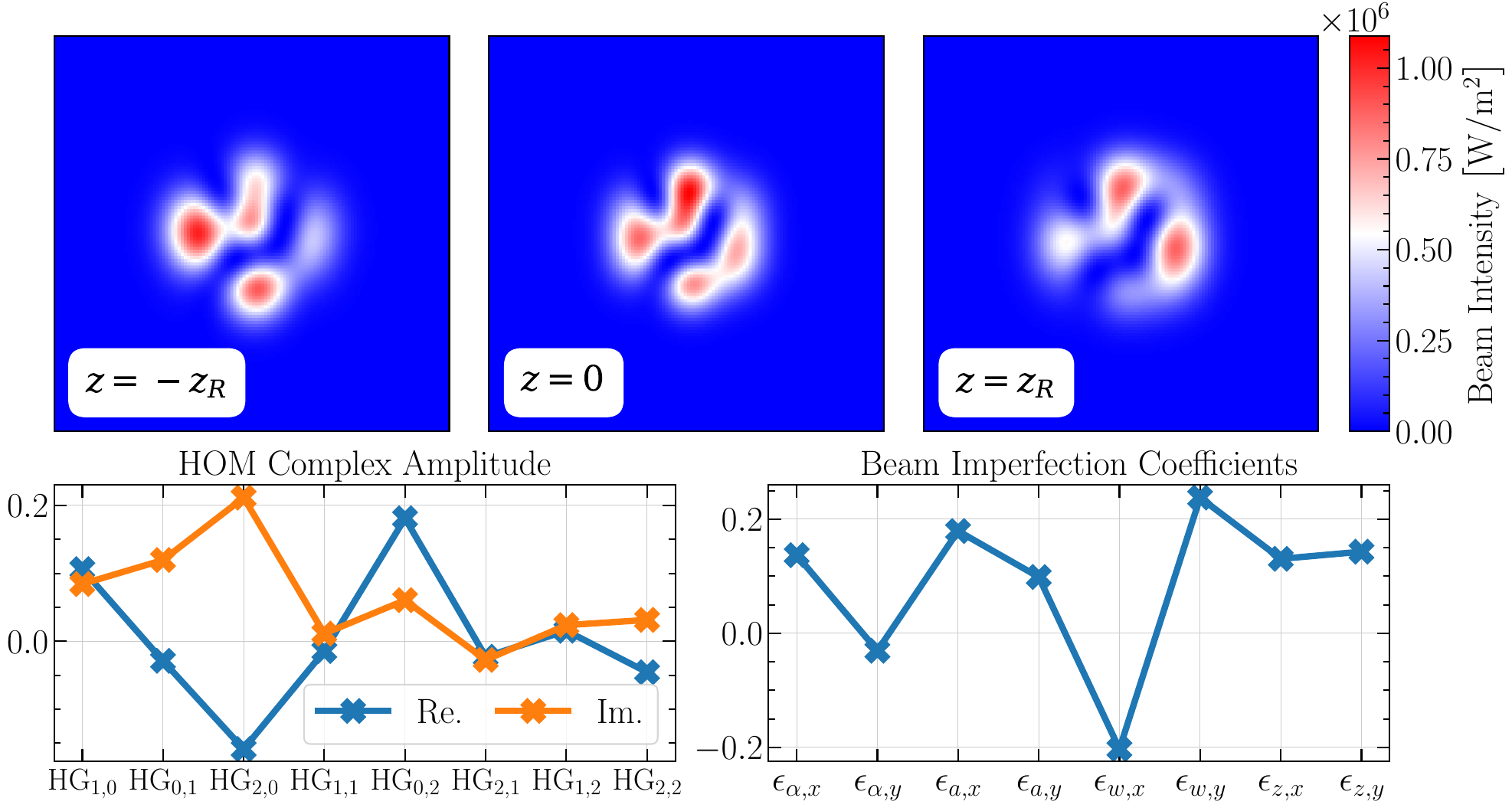}
    \caption{Example input for the ML pipeline. \textit{Top}: intensity images captured at three different locations along the beam propagation direction. \textit{Bottom left}: underlying HOM coefficients, split into real and imaginary parts. \textit{Bottom right}: beam imperfection parameters for all eight misalignment and mode mismatch degrees of freedom.}
    \label{fig-beam_intensity_3z}
\end{figure}

Fig.~\ref{fig-beam_intensity_3z} shows an example data sample. The top row shows the intensity images captured at three different locations along the beam propagation direction. Variations in shape and intensity along the propagation direction encode the phase profile and the modal content of the perturbed beam. In the bottom row, the left and right panels display the corresponding complex HOM contents and the underlying beam imperfection parameters, respectively. The machine learning pipeline aims to take the three intensity images in the top row, easily and rapidly obtained with CCD cameras, to extract the HOM contents (bottom left panel) using the first CNN model, and then use these predicted HOM coefficients as input to the second DL model to predict the imperfection parameters for the complete set of misalignment and mode mismatch (bottom right panel).

\subsection{Model Architecture}
For the task of multi-intensity-image-based mode decomposition, we adopt a pre-trained convolutional neural network, specifically the VGG16 architecture~\cite{VGG16} implemented in TensorFlow Keras. The top classification layers are removed, and custom fully connected layers are appended to perform regression on the real and imaginary parts of the complex mode coefficients. As illustrated in Fig.~\ref{fig-concept_fig2}, the three intensity images are stacked along the channel dimension, analogous to RGB channels, and fed into the network. For unit input power, the intensity images have peak values of approximately \(10^6\, \mathrm{W/m^2}\) (see Fig.~\ref{fig-beam_intensity_3z}). To improve numerical stability and training efficiency, the intensity images are normalized by dividing by the maximum intensity value.

The VGG16 backbone is followed by two dense layers with 320 and 256 neurons, respectively, each using ReLU activation. The output layer consists of 16 neurons corresponding to the real and imaginary parts of the HOM coefficients for modes with $n,m \leq 2$. Since the label range for these coefficients is $[-1/4, 1/4]$ (see Tab.~\ref{tab-data_param}), we scale the labels to $[0,1]$ using the \textit{MinMaxScaler} from \textit{scikit-learn}. Accordingly, the output layer uses a Sigmoid activation function to ensure predictions lie within this range.

For the second deep learning model, the input consists of the predicted HOM coefficients (16 neurons) along with one additional neuron for the amplitude of the remaining fundamental mode, resulting in a total of 17 input neurons. This input passes through three fully connected layers, each containing 512 neurons with ReLU activation. The final regression layer employs a linear activation function and outputs eight parameters corresponding to the entire set of degrees of freedom associated with misalignment and mode mismatch.

\subsection{Training and Testing}

For the first CNN model in the pipeline, the VGG16 network was initialized with weights pre-trained on ImageNet, which contains approximately 1.2 million images across 1000 categories~\cite{ImageNet}. The CNN network is subsequently trained on 80,000 beam-intensity samples. The Adam optimizer was used with an initial learning rate of $1\times10^{-5}$ and a batch size of 32~\cite{kingma2017adammethodstochasticoptimization}. To improve convergence and prevent overfitting, an adaptive learning rate schedule was implemented, where the learning rate was halved if the validation loss failed to improve for three consecutive epochs.

We adopted a multi-stage progressive training strategy, splitting the 80,000 samples into four subsets of 20,000 samples each. The model was first trained on the initial subset, and the best performing model from this stage was used as the starting point for subsequent stages. Each stage refines the previously learned model, allowing incremental learning starting from a model already near a good local minimum. The mean squared error (MSE) was used as the loss function.

\begin{figure}[t]
    \centering
    \includegraphics[width=0.7\linewidth]{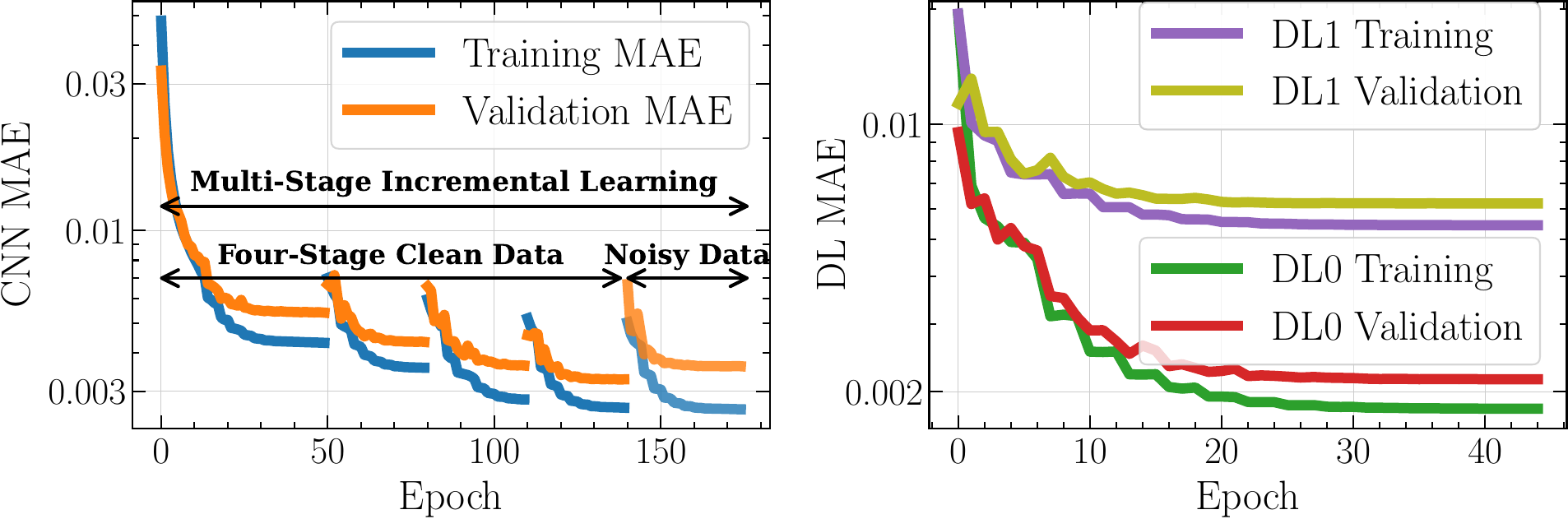}
    \caption{Training history for the machine learning pipeline. \textit{Left}: The CNN model trained using a multi-stage incremental learning strategy, consisting of four stages with clean data and one stage with noisy data. \textit{Right}: The DL model trained either on the ground truth HOM coefficients (DL0) or on the HOM coefficients predicted by the upstream CNN model (DL1).}
    \label{fig-learning_hist_CNN}
\end{figure}

The first four line segments in the left panel of Fig.~\ref{fig-learning_hist_CNN} illustrate the training history for this four-stage incremental learning process. We use the mean absolute error (MAE) as a performance evaluation metric, defined as
\begin{equation}
\begin{aligned}
\mathrm{MAE} &= 
\frac{1}{16} \sum_{n,m} 
\left(
\left| \Delta \operatorname{Re}\!\left(c_{n,m}\right) \right|
+ 
\left| \Delta \operatorname{Im}\!\left(c_{n,m}\right) \right|
\right), \\
&\text{where} \quad 
\Delta c_{n,m} = c_{n,m}^{\prime} - c_{n,m} ,
\end{aligned}
\end{equation}
i.e., the difference between the predicted coefficients and the ground truth, which in total includes sixteen real and imaginary components corresponding to eight higher-order modes. As shown in Fig.~\ref{fig-learning_hist_CNN}, the four-stage incremental learning approach progressively improves the model's performance on both the training and validation sets. Specifically, the MAE for the best model prediction on the validation set after each stage of training is 0.0054, 0.0043, 0.0036, and 0.0033, demonstrating consistent improvement in predictive accuracy.

This four-stage incremental training is based entirely on clean samples, with no noise added to the input intensity images. However, this does not represent realistic experimental conditions, as detector noise is ubiquitous in optics measurements. To evaluate the robustness of the best-trained CNN model against measurement noise, we introduce random fluctuations into the intensity inputs. These fluctuations are modeled as Gaussian noise, simulating thermal noise in electronics or readout noise in CCD cameras~\cite{Mannam:22}. Specifically, we add Gaussian noise to each pixel of the CCD intensity image as
\begin{equation}
I_{\text {noisy}}(x, y)=\max \left[I_{\text {0}}(x, y)+\eta I_{max} \cdot \mathcal{N}(0,1), \,0\right] ,
\end{equation}
where $\mathcal{N}(0,1)$ is the standard Gaussian distribution, $I_{0}(x, y)$ is the intensity without noise (i.e., the ``clean'' sample), and $I_{\max}$ is the maximum intensity. The \textit{max} operation ensures that all pixel values remain non-negative. The parameter $\eta \in [0, 1]$ quantifies the noise level as a fraction of the maximum intensity $I_{\max}$.

\begin{figure}[t]
    \centering
    \includegraphics[width=0.6\linewidth]{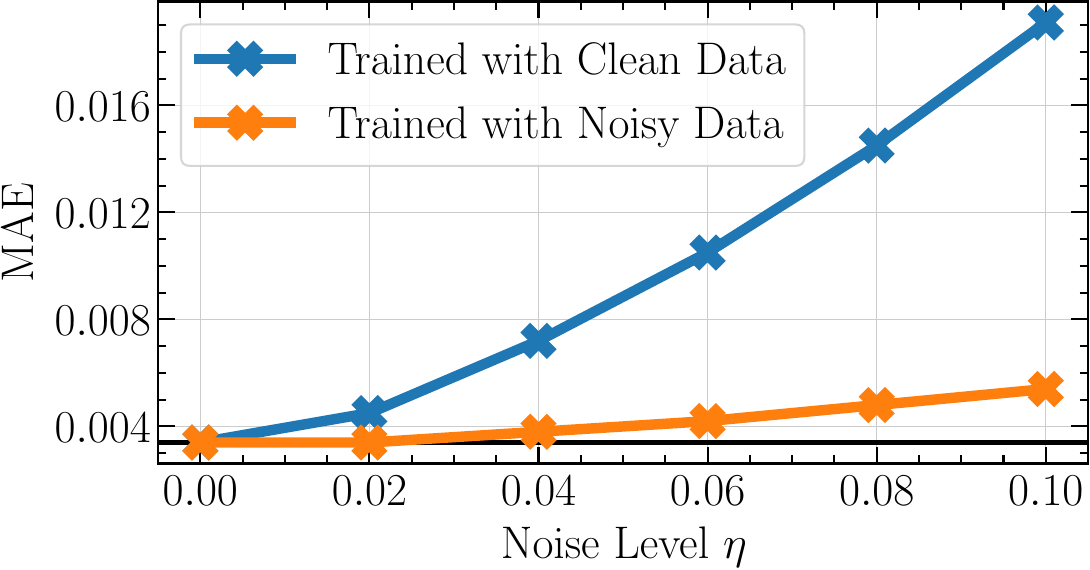}
    \caption{The performance of the CNN model in terms of MAE on a test set of 1000 samples with increasing noise levels. Blue markers show results for the model trained only on clean data, while orange markers show results for the model retrained with noisy data using incremental learning.}
    \label{fig-MAE_clean_noisy}
\end{figure}

We gradually increase the noise level in the test dataset, and evaluate the performance of the model that was trained on clean data. The results are shown in Fig.~\ref{fig-MAE_clean_noisy} as blue crosses. Noise levels are increased up to 0.1. As shown, the MAE on the test set increases significantly from 0.0034 when the noise level is zero to a maximum of around 0.019.

To address this degradation in predictability in the presence of noise, we introduce an additional stage in the incremental training where the model is continued to be trained with the same type of Gaussian noise contamination applied to the training data. This allows the model to adapt to noisy conditions by adjusting its learned weights for robustness. The learning history for this noisy-data training stage over another 20,000 samples is shown as the final line segments in the left panel of Fig.~\ref{fig-learning_hist_CNN}, where the MAE decreases as the model learns from noisy inputs.

We then reevaluate the model's performance against a noisy test set with increasing noise levels. The results, shown as the orange crosses in Fig.~\ref{fig-MAE_clean_noisy}, demonstrate that the MAE increases only slightly, reaching a maximum of around 0.0054 at a 10\% noise level. This value is much closer to the optimal predictability evaluated on clean test data, demonstrating that the retrained model achieves near-optimal performance even on significant noisy inputs.

\begin{figure}[t]
    \centering
    \includegraphics[width=0.6\linewidth]{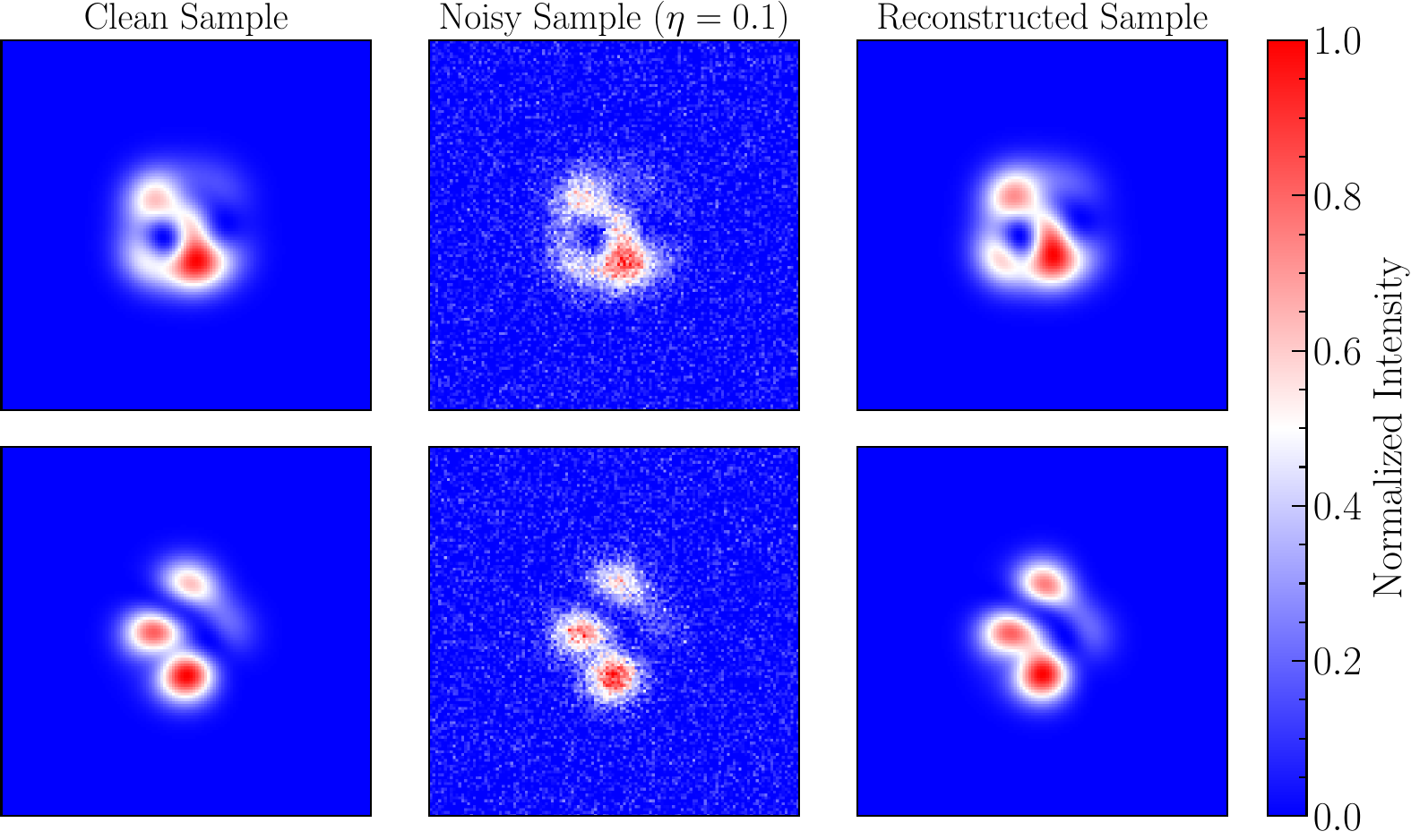}
    \caption{Two example noisy sample images. Each row shows the clean image (left), the image with $\eta=0.1$ Gaussian intensity noise (middle), and the reconstructed image based on the predicted HOM coefficients (right).}
    \label{fig-HOM_CNN_denoise_image}
\end{figure}

Fig.~\ref{fig-HOM_CNN_denoise_image} shows two examples of intensity images with random noise: the clean sample (left), the noisy sample with $\eta=0.1$ (middle), and the reconstructed sample from the CNN model using the predicted complex mode coefficients with noisy input (right). The results demonstrate that our model is robust to Gaussian noise in the intensity images, effectively preserving the original intensity structure. This denoising capability of the CNN model also suggests potential applications beyond beam diagnostics, such as general image denoising and processing tasks~\cite{Mannam:22}.

For the second stage of the ML pipeline, where we take the predicted HOM coefficients from the upstream CNN model along with the amplitude of the remaining fundamental mode as input and output the complete set of misalignment and mode mismatch parameters, we similarly employ the Adam optimizer with an initial learning rate of $1\times 10^{-3}$ and a batch size of $32$. The same adaptive learning rate scheduler is adopted to improve convergence. For completeness, we train this DL model in two cases. First, we train it using the actual HOM coefficients from the input dataset (the ground truth) to establish a theoretical performance limit, assuming perfect knowledge of the HOM contents. This model is referred to as ``DL0''. Second, we train the model using the coefficients predicted by the previous CNN model as input. This model, referred to as ``DL1'', captures the residual error introduced by the first stage of the mode decomposition and thus reflects the total residual error for the final prediction of beam imperfection parameters.

The training process for both DL models is shown in the right panel of Fig.~\ref{fig-learning_hist_CNN}. Both models exhibit steady improvement with increasing epochs and plateau at approximately $30$ epochs. The models are trained on $20,000$ samples, with no significant enhancement in predictability observed at this stage via incremental learning.

The DL0 model, trained on the ground truth HOM coefficients, achieves a minimum MAE of 0.0021 on the validation set. In contrast, the DL1 model, trained on the HOM coefficients predicted by the upstream CNN model, reaches a minimum MAE of 0.0062. This difference is expected, as the CNN model carries a residual MAE error of 0.0034 on the validation and test sets, as shown in the left panel of Fig.~\ref{fig-learning_hist_CNN} and in Fig.~\ref{fig-MAE_clean_noisy}. The downstream DL1 model inherits this residual error, resulting in a total error of 0.0062 for the complete two-step pipeline.

\section{Result and Discussion \label{sec-result}}

With the trained machine learning pipeline, we demonstrate the ability to perform full beam alignment and mode mismatch sensing from intensity-only measurements. 

\begin{figure}[t]
    \centering
    \includegraphics[width=0.7\linewidth]{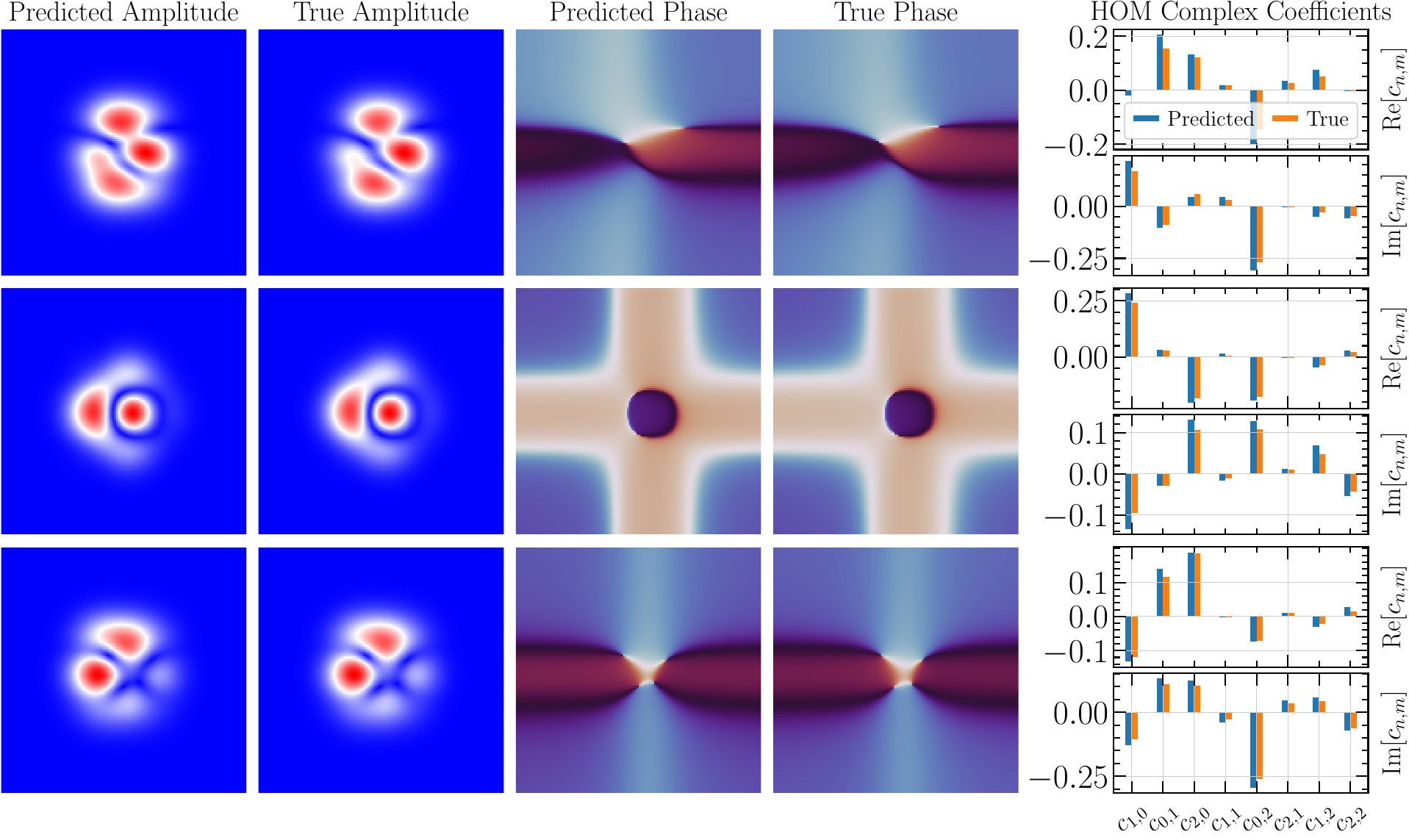}
    \caption{Prediction results of the intensity-based mode decomposition model using a CNN. Three example cases are shown in three rows. The five columns display: (1) the predicted mode amplitude reconstructed from the predicted mode coefficients, (2) the actual mode amplitude, (3) the predicted phase, (4) the actual phase, and (5) a comparison of predicted versus true mode amplitudes.}
    \label{fig-complex_intensity_ML_output}
\end{figure}

Fig.~\ref{fig-complex_intensity_ML_output} shows example prediction results for the intensity-based mode decomposition model. It presents the predicted beam amplitude and phase profiles, compared against their ground truth, as well as a direct comparison of the corresponding complex mode coefficients, split into real and imaginary parts. As shown, the CNN-based mode decomposition predicts the mode coefficients with high precision, achieving an MAE of 0.0034 over the test set. This ensures an accurate reconstruction of the beam's complex amplitude using intensity only measurements at multiple propagation locations. Importantly, as shown, using multiple intensity images resolves the phase ambiguity inherent to single-intensity-image methods. In our setup, three intensity images, measured at the beam waist and one Rayleigh range before and after the waist, provide sufficient phase diversity to reconstruct the complex field amplitudes with high fidelity.

This novel mode decomposition technique can be extended to include different sets of higher-order modes tailored for various beam diagnostics applications, such as studies of spherical aberrations or active wavefront control in higher-power interferometry. To do so, one only needs to generate the appropriate labeled dataset containing a random source of aberrations and collect the resulting beam intensity images together with the corresponding complex mode amplitudes, serving as the input and output, respectively, similar to the workflow presented in this paper. 

\begin{figure*}[t]
    \centering
    \includegraphics[width=0.7\linewidth]{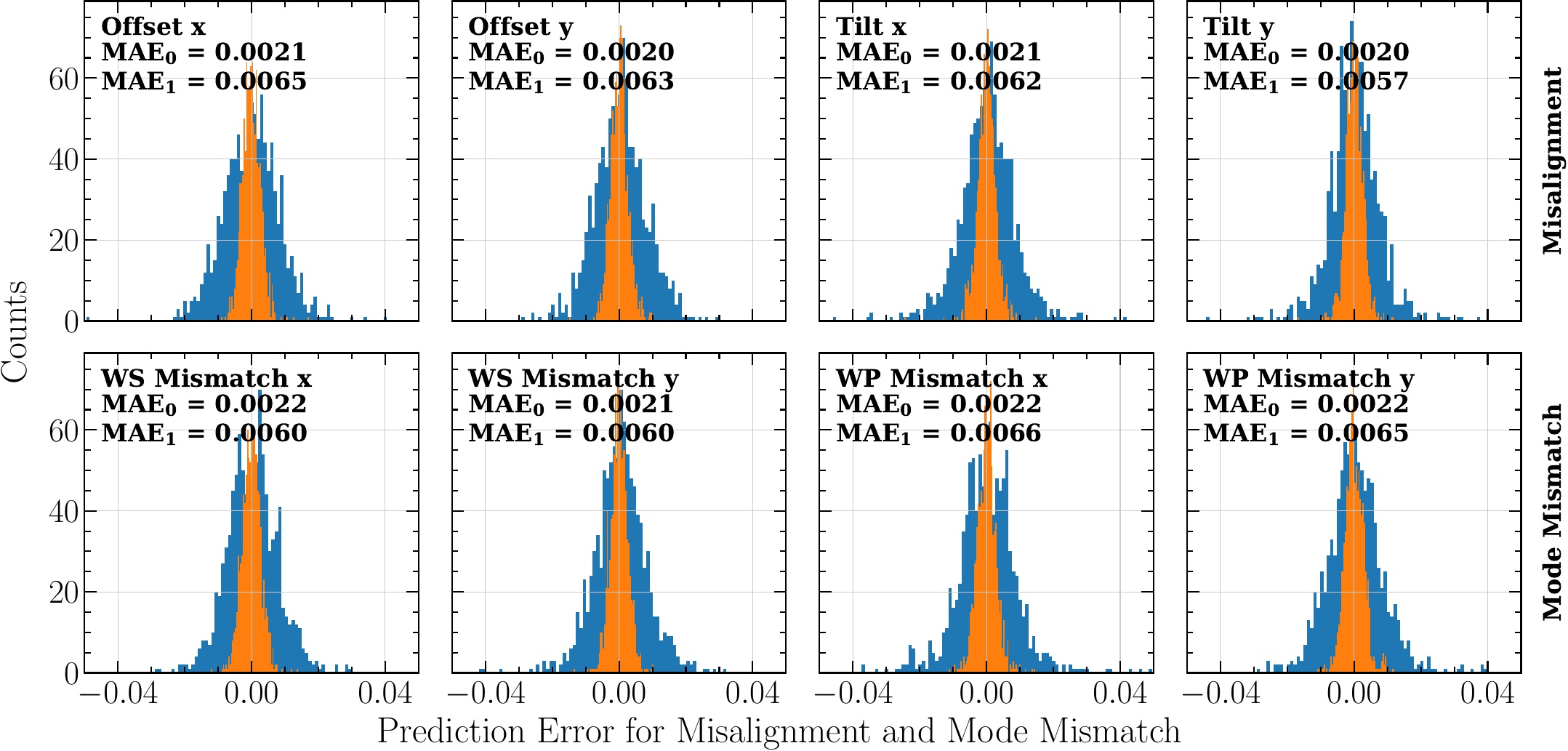}
    \caption{Prediction results for misalignment and mode mismatch expressed as absolute errors across all eight DoFs, shown as histograms for a test set of 1000 samples. The top row corresponds to the misalignment DoFs, and the bottom row corresponds to the mode mismatch DoFs. The MAE for each DoF is indicated in the respective panel. MAE$_0$ (orange) uses ground truth complex coefficients, whereas MAE$_1$ (blue) uses predicted coefficients from the CNN as input.}
    \label{fig-Misalignment_MM_ML_predict_error}
\end{figure*}

The output of the CNN-based mode decomposition serves as the input to the downstream regression model, which predicts the full set of beam imperfection parameters. Fig.~\ref{fig-Misalignment_MM_ML_predict_error} presents histograms of the prediction errors for all eight DoFs over the test set, defined as the difference between the true and predicted imperfections. The top row shows the four orthogonal misalignment DoFs, while the bottom row shows the four mode mismatch DoFs. The two cases defined in \S\ref{sec-setup} are presented: (i) the model prediction error using the ground truth HOM coefficients as input (DL0), shown in orange, and (ii) the model prediction error using the predicted HOM coefficients from the upstream CNN model (DL1), shown in blue. The MAE is displayed at the top of each panel, with ``MAE$_0$'' and ``MAE$_1$'' denoting the errors for DL0 and DL1, respectively. A few rare outliers exist for DL1 with 0.05 $<$ MAE $<$ 0.15, which can be mitigated through complementary measurements, such as cross-validation with existing wavefront sensing techniques, or consistency checks across successive measurements.

We achieve nearly uniform prediction accuracy across all eight imperfection coefficients, indicating that the model effectively learns the dynamics of beam misalignment and mode mismatch with comparable performance. The average MAE across all eight outputs is 0.0021 for DL0 and 0.0062 for DL1. The difference between these values is primarily attributable to the residual prediction error propagated from the upstream CNN model, as explained in \S\ref{sec-setup}.

For instance, for tilt in the $X$ direction, the final MAE from the two-step machine learning pipeline is 0.0062. This corresponds to a residual tilt error of
\begin{equation}
    \Delta \alpha = \Delta \epsilon_{\alpha, x} \cdot \Theta = 0.0062 \cdot \Theta=4.2\, \mu \text{rad} ,
\end{equation}
where $\Theta$ is the beam far-field divergence angle. The residual power loss from the tilt would be
\begin{equation}
    \text{Residual Power Loss} = (\Delta \epsilon_{\alpha, x})^2 =  38\, \text{ppm} .
\end{equation}

\begin{table}[b]
\centering
\caption{Summary of MAE, the physical residual error, and the corresponding residual power loss for all eight misalignment and mode mismatch degrees of freedom on the test set. The total power loss over all beam imperfection parameters is 310~ppm.}
\setlength{\tabcolsep}{10pt}
\begin{tabular}{c|c|c|c}
\hline
\hline
Parameter & \hspace{2pt} MAE \hspace{2pt} & Residual Error & Residual Loss \\
\hline
$a_{x}$ & 0.0065 & 3.2 $\mu m$ & 42 ppm \\ 
$a_{y}$ & 0.0063 & 3.2 $\mu m$ & 40 ppm \\
$\alpha_{x}$ & 0.0062 & 4.2 $\mu rad$ & 38 ppm \\
$\alpha_{y}$ & 0.0057 & 3.9 $\mu rad$ & 33 ppm \\
$w_{x}$ & 0.0060 & 4.2 $\mu m$ & 36 ppm \\
$w_{y}$ & 0.0060 & 4.3 $\mu m$ & 36 ppm \\
$z_{x}$ & 0.0066 & 1.4 $cm$ & 43 ppm \\
$z_{y}$ & 0.0065 & 1.4 $cm$ & 42 ppm \\
\hline
\hline
\end{tabular}
\label{tab-result_param}
\end{table}

The residual error and power loss for all eight imperfection parameters are summarized in Tab.~\ref{tab-result_param}. The total residual power loss for \textit{all} eight misalignment and mode mismatch DoFs sums to 310~ppm. This is significantly smaller than the typical mode mismatch-induced coupling losses in the coupled-cavity system in current gravitational-wave detectors, which are typically on the order of a few percent~\cite{PhysRevD.111.062002}.

\section{Conclusion \label{sec-conclusion}}
In this work, we presented a novel two-step machine learning framework for multi-intensity-based mode decomposition and subsequent \textit{simultaneous} beam misalignment and mode mismatch sensing in optical cavities. Our approach uses a convolutional neural network adapted from the VGG16 architecture to infer the complex Hermite-Gaussian mode coefficients directly from a series of three intensity images, without requiring complex interferometric sensing signals or phase measurements. The predicted complex mode coefficients are then used as input to a downstream neural network that \textit{simultaneously} predicts all eight degrees of freedom associated with generic misalignment and mode mismatch.

We demonstrated that the proposed ML pipeline can accurately and robustly reconstruct the modal content of perturbed beams and infer the underlying misalignment and mode mismatch states. The CNN-based mode decomposition model achieves a mean absolute error of approximately 0.0034 for the complex mode coefficients for $\mathrm{HG}_{n,m}$ modes with the indices $n,m \leq2$. The entire pipeline, as a result, achieves an MAE of approximately 0.0062 across all degrees of freedom, corresponding to a total residual optical loss below 310~ppm, significantly lower than the optical loss expected in current gravitational-wave detectors. These results highlight the potential of ML-based techniques to provide precise sensing of beam alignment and mode-matching states using only simplistic beam intensity imaging diagnostics.

The presented method offers a pathway toward implementing camera-based, real-time beam characterization in high-precision interferometers, where traditional phase-camera systems are limited by hardware complexity and systematic uncertainties. Future extensions of this work will involve experimental validation of the proposed technique for in situ cavity diagnostics. Realistic imperfections relevant to practical implementations, such as mirror surface roughness, scattering losses, ghost beams, and beam clipping, are not included in the present model. Assessing the robustness of the method under such non-ideal conditions will be an important step toward in situ application. Additional efforts will explore applications to extend the ML framework to capture higher-order aberrations, beyond the linear and quadratic terms in the wavefront distortions associated with misalignment and mode mismatch. Time-dependent perturbations, such as thermally induced beam wavefront distortions under high intracavity power in gravitational-wave detectors, introduce additional complexity that may require adaptive or online learning strategies, where the model is continuously updated as the thermal state of the test mass evolves to ensure robustness and accuracy. Ultimately, the integration of ML-based optical mode analysis and wavefront diagnostics could enable continuous, high-fidelity monitoring of beam quality and cavity mode matching, providing critical feedback for thermal compensation and the adaptive mode sensing and control systems in next-generation gravitational-wave detectors.

\begin{backmatter}

\bmsection{Funding}
This work has received support from the investment programme ``France 2030'' as part of the IdEx programme (ANR-18-IDEX-0001) implemented by Universit\'e Paris Cit\'e, under which the inIdEx project QuanTech@Paris is conducted; from Quantum-FRESCO project (ANR-23-CE31-0004) and from the Paris Île-de-France Region.

\bmsection{Acknowledgment}
The authors thank Antonio Perreca for helpful comments during the preparation of this manuscript. This document was submitted to Virgo and LIGO collaborations with the numbers VIR-0183A-26 and P2600081.

\bmsection{Disclosures}
The authors declare no conflicts of interest.

\bmsection{Data Availability Statement}
Data underlying the results presented in this paper are not publicly available at this time but may be obtained from the authors upon reasonable request.

\end{backmatter}

%%%%%%%%%% If using BibTeX:
\bibliography{sample}

@article{Vitale_2021,
   title={The first 5 years of gravitational-wave astrophysics},
   volume={372},
   ISSN={1095-9203},
   url={http://dx.doi.org/10.1126/science.abc7397},
   DOI={10.1126/science.abc7397},
   number={6546},
   journal={Science},
   publisher={American Association for the Advancement of Science (AAAS)},
   author={Vitale, Salvatore},
   year={2021},
   month=jun }

@misc{GWTC-4,
      title={GWTC-4.0: Updating the Gravitational-Wave Transient Catalog with Observations from the First Part of the Fourth LIGO-Virgo-KAGRA Observing Run}, 
      author={Abbott \emph{et al.}, R.},
      collaboration = {LIGO Scientific Collaboration, Virgo Collaboration, and KAGRA Collaboration},
      year={2025},
      eprint={2508.18082},
      archivePrefix={arXiv},
      primaryClass={gr-qc},
      url={https://arxiv.org/abs/2508.18082}, 
}

@techreport{CEHorizonStudy,
    title={{A Horizon Study for Cosmic Explorer: Science, Observatories, and Community}}, 
    author={M. Evans and R. X. Adhikari and C. Afle and S. W. Ballmer and S. Biscoveanu and S. Borhanian and D. A. Brown and Y. Chen and R. Eisenstein and A. Gruson and A. Gupta and E. D. Hall and R. Huxford and B. Kamai and R. Kashyap and J. S. Kissel and K. Kuns and P. Landry and A. Lenon and G. Lovelace and L. McCuller and K. Ng and A. H. Nitz and J. Read and B. S. Sathyaprakash and D. H. Shoemaker and B. Slagmolen and J. R. Smith and V. Srivastava and L. Sun and S. Vitale and R. Weiss},
    year={2021},
    month = oct,
    Type = {arXiv e-Print},
    Number = {2109.09882},
    eprint={2109.09882},
    archivePrefix={arXiv},
    primaryClass={astro-ph.IM},
    eid = {arXiv:2109.09882},
    url = {https://doi.org/10.48550/arXiv.2109.09882}
}

@article{Caves:1980,
    title = {{On the measurement of a weak classical force coupled to a quantum-mechanical oscillator. I. Issues of principle}},
    author = {Caves, Carlton M. and Thorne, Kip S. and Drever, Ronald W. P. and Sandberg, Vernon D. and Zimmermann, Mark},
    journal = {Rev. Mod. Phys.},
    volume = {52},
    issue = {2},
    pages = {341--392},
    numpages = {0},
    year = {1980},
    month = {Apr},
    publisher = {American Physical Society},
    doi = {10.1103/RevModPhys.52.341},
    url = {https://link.aps.org/doi/10.1103/RevModPhys.52.341}
}

@article{Caves:1981,
    title = {{Quantum-mechanical noise in an interferometer}},
    author = {Caves, Carlton M.},
    journal = {Phys. Rev. D},
    volume = {23},
    issue = {8},
    pages = {1693--1708},
    numpages = {0},
    year = {1981},
    month = {Apr},
    publisher = {American Physical Society},
    doi = {10.1103/PhysRevD.23.1693},
    url = {https://link.aps.org/doi/10.1103/PhysRevD.23.1693}
}

@article{ET_2010,
doi = {10.1088/0264-9381/27/19/194002},
url = {https://doi.org/10.1088/0264-9381/27/19/194002},
year = {2010},
month = {sep},
publisher = {},
volume = {27},
number = {19},
pages = {194002},
author = {Punturo, M and Abernathy, M and Acernese, F and Allen, B and Andersson, N and Arun, K and Barone, F and Barr, B and Barsuglia, M and Beker, M and Beveridge, N and Birindelli, S and Bose, S and Bosi, L and Braccini, S and Bradaschia, C and Bulik, T and Calloni, E and Cella, G and Mottin, E Chassande and Chelkowski, S and Chincarini, A and Clark, J and Coccia, E and Colacino, C and Colas, J and Cumming, A and Cunningham, L and Cuoco, E and Danilishin, S and Danzmann, K and De Luca, G and De Salvo, R and Dent, T and De Rosa, R and Di Fiore, L and Di Virgilio, A and Doets, M and Fafone, V and Falferi, P and Flaminio, R and Franc, J and Frasconi, F and Freise, A and Fulda, P and Gair, J and Gemme, G and Gennai, A and Giazotto, A and Glampedakis, K and Granata, M and Grote, H and Guidi, G and Hammond, G and Hannam, M and Harms, J and Heinert, D and Hendry, M and Heng, I and Hennes, E and Hild, S and Hough, J and Husa, S and Huttner, S and Jones, G and Khalili, F and Kokeyama, K and Kokkotas, K and Krishnan, B and Lorenzini, M and Lück, H and Majorana, E and Mandel, I and Mandic, V and Martin, I and Michel, C and Minenkov, Y and Morgado, N and Mosca, S and Mours, B and Müller–Ebhardt, H and Murray, P and Nawrodt, R and Nelson, J and Oshaughnessy, R and Ott, C D and Palomba, C and Paoli, A and Parguez, G and Pasqualetti, A and Passaquieti, R and Passuello, D and Pinard, L and Poggiani, R and Popolizio, P and Prato, M and Puppo, P and Rabeling, D and Rapagnani, P and Read, J and Regimbau, T and Rehbein, H and Reid, S and Rezzolla, L and Ricci, F and Richard, F and Rocchi, A and Rowan, S and Rüdiger, A and Sassolas, B and Sathyaprakash, B and Schnabel, R and Schwarz, C and Seidel, P and Sintes, A and Somiya, K and Speirits, F and Strain, K and Strigin, S and Sutton, P and Tarabrin, S and Thüring, A and van den Brand, J and van Leewen, C and van Veggel, M and van den Broeck, C and Vecchio, A and Veitch, J and Vetrano, F and Vicere, A and Vyatchanin, S and Willke, B and Woan, G and Wolfango, P and Yamamoto, K},
title = {The Einstein Telescope: a third-generation gravitational wave observatory},
journal = {Classical and Quantum Gravity},
}

@techreport{LIGOwhitepaper2025,
    author = {{LIGO Scientific Collaboration}},
    title = {{The LSC Instrument Science White Paper (2025 edition)}},
    Type = {LIGO Technical Report},
    Number = {LIGO-T2400407},
    Year = {2024},
    Month = dec,
    url = {https://dcc.ligo.org/LIGO-T2400407/public}
}

@techreport{PostO5Report:2022,
    Author = {P. Fritschel and K. Kuns and J. Driggers and A. Effler and B. Lantz and D. Ottaway and S. Ballmer and K. Dooley and R. X. Adhikari and M. Evans and B. Farr and G. Gonzalez and P. Schmidt and S. Raja},
    Title = {{Report of the LSC Post-O5 Study Group}},
    Type = {LIGO Technical Report},
    Number = {LIGO-T2200287},
    Year = {2022},
    Month = nov,
    url = {https://dcc.ligo.org/LIGO-T2200287/public}
}

@article{Kwee,
  title = {Decoherence and degradation of squeezed states in quantum filter cavities},
  author = {Kwee, P. and Miller, J. and Isogai, T. and Barsotti, L. and Evans, M.},
  journal = {Phys. Rev. D},
  volume = {90},
  issue = {6},
  pages = {062006},
  numpages = {12},
  year = {2014},
  month = {Sep},
  publisher = {American Physical Society},
  doi = {10.1103/PhysRevD.90.062006},
  url = {https://link.aps.org/doi/10.1103/PhysRevD.90.062006}
}

@article{FDS_MIT_2020,
  title = {Frequency-Dependent Squeezing for Advanced LIGO},
  author = {McCuller, L. and Whittle, C. and Ganapathy, D. and Komori, K. and Tse, M. and Fernandez-Galiana, A. and Barsotti, L. and Fritschel, P. and MacInnis, M. and Matichard, F. and Mason, K. and Mavalvala, N. and Mittleman, R. and Yu, Haocun and Zucker, M. E. and Evans, M.},
  journal = {Phys. Rev. Lett.},
  volume = {124},
  issue = {17},
  pages = {171102},
  numpages = {5},
  year = {2020},
  month = {Apr},
  publisher = {American Physical Society},
  doi = {10.1103/PhysRevLett.124.171102},
  url = {https://link.aps.org/doi/10.1103/PhysRevLett.124.171102}
}

@article{Ding25,
  title = {Performance of multiple filter-cavity schemes for frequency-dependent squeezing in gravitational-wave detectors},
  author = {Ding, Jacques and Capocasa, Eleonora and Ahrend, Isander and Liu, Fangfei and Zhao, Yuhang and Barsuglia, Matteo},
  journal = {Phys. Rev. D},
  volume = {112},
  issue = {12},
  pages = {122001},
  numpages = {21},
  year = {2025},
  month = {Dec},
  publisher = {American Physical Society},
  doi = {10.1103/4s4d-9mb4},
  url = {https://link.aps.org/doi/10.1103/4s4d-9mb4}
}

@article{yuhangzhao_2020,
  title = {Frequency-Dependent Squeezed Vacuum Source for Broadband Quantum Noise Reduction in Advanced Gravitational-Wave Detectors},
  author = {Zhao, Yuhang and Aritomi, Naoki and Capocasa, Eleonora and Leonardi, Matteo and Eisenmann, Marc and Guo, Yuefan and Polini, Eleonora and Tomura, Akihiro and Arai, Koji and Aso, Yoichi and Huang, Yao-Chin and Lee, Ray-Kuang and L\"uck, Harald and Miyakawa, Osamu and Prat, Pierre and Shoda, Ayaka and Tacca, Matteo and Takahashi, Ryutaro and Vahlbruch, Henning and Vardaro, Marco and Wu, Chien-Ming and Barsuglia, Matteo and Flaminio, Raffaele},
  journal = {Phys. Rev. Lett.},
  volume = {124},
  issue = {17},
  pages = {171101},
  numpages = {7},
  year = {2020},
  month = {Apr},
  publisher = {American Physical Society},
  doi = {10.1103/PhysRevLett.124.171101},
  url = {https://link.aps.org/doi/10.1103/PhysRevLett.124.171101}
}

@article{Ganapathy_2023,
  title = {Broadband Quantum Enhancement of the LIGO Detectors with Frequency-Dependent Squeezing},
  author = {Ganapathy, D. and Jia, W. and Nakano, M. and Xu, V. and Aritomi, N. and Cullen, T. and Kijbunchoo, N. and Dwyer, S. E. and Mullavey, A. and McCuller, L. and Abbott, R. and Abouelfettouh, I. and Adhikari, R. X. and Ananyeva, A. and Appert, S. and Arai, K. and Aston, S. M. and Ball, M. and Ballmer, S. W. and Barker, D. and Barsotti, L. and Berger, B. K. and Betzwieser, J. and Bhattacharjee, D. and Billingsley, G. and Biscans, S. and Bode, N. and Bonilla, E. and Bossilkov, V. and Branch, A. and Brooks, A. F. and Brown, D. D. and Bryant, J. and Cahillane, C. and Cao, H. and Capote, E. and Clara, F. and Collins, J. and Compton, C. M. and Cottingham, R. and Coyne, D. C. and Crouch, R. and Csizmazia, J. and Dartez, L. P. and Demos, N. and Dohmen, E. and Driggers, J. C. and Effler, A. and Ejlli, A. and Etzel, T. and Evans, M. and Feicht, J. and Frey, R. and Frischhertz, W. and Fritschel, P. and Frolov, V. V. and Fulda, P. and Fyffe, M. and Gateley, B. and Giaime, J. A. and Giardina, K. D. and Glanzer, J. and Goetz, E. and Goetz, R. and Goodwin-Jones, A. W. and Gras, S. and Gray, C. and Griffith, D. and Grote, H. and Guidry, T. and Hall, E. D. and Hanks, J. and Hanson, J. and Heintze, M. C. and Helmling-Cornell, A. F. and Holland, N. A. and Hoyland, D. and Huang, H. Y. and Inoue, Y. and James, A. L. and Jennings, A. and Karat, S. and Karki, S. and Kasprzack, M. and Kawabe, K. and King, P. J. and Kissel, J. S. and Komori, K. and Kontos, A. and Kumar, R. and Kuns, K. and Landry, M. and Lantz, B. and Laxen, M. and Lee, K. and Lesovsky, M. and Llamas, F. and Lormand, M. and Loughlin, H. A. and Macas, R. and MacInnis, M. and Makarem, C. N. and Mannix, B. and Mansell, G. L. and Martin, R. M. and Mason, K. and Matichard, F. and Mavalvala, N. and Maxwell, N. and McCarrol, G. and McCarthy, R. and McClelland, D. E. and McCormick, S. and McRae, T. and Mera, F. and Merilh, E. L. and Meylahn, F. and Mittleman, R. and Moraru, D. and Moreno, G. and Nelson, T. J. N. and Neunzert, A. and Notte, J. and Oberling, J. and O'Hanlon, T. and Osthelder, C. and Ottaway, D. J. and Overmier, H. and Parker, W. and Pele, A. and Pham, H. and Pirello, M. and Quetschke, V. and Ramirez, K. E. and Reyes, J. and Richardson, J. W. and Robinson, M. and Rollins, J. G. and Romel, C. L. and Romie, J. H. and Ross, M. P. and Ryan, K. and Sadecki, T. and Sanchez, A. and Sanchez, E. J. and Sanchez, L. E. and Savage, R. L. and Schaetzl, D. and Schiworski, M. G. and Schnabel, R. and Schofield, R. M. S. and Schwartz, E. and Sellers, D. and Shaffer, T. and Short, R. W. and Sigg, D. and Slagmolen, B. J. J. and Soike, C. and Soni, S. and Srivastava, V. and Sun, L. and Tanner, D. B. and Thomas, M. and Thomas, P. and Thorne, K. A. and Torrie, C. I. and Traylor, G. and Ubhi, A. S. and Vajente, G. and Vanosky, J. and Vecchio, A. and Veitch, P. J. and Vibhute, A. M. and von Reis, E. R. G. and Warner, J. and Weaver, B. and Weiss, R. and Whittle, C. and Willke, B. and Wipf, C. C. and Yamamoto, H. and Zhang, L. and Zucker, M. E.},
  collaboration = {LIGO O4 Detector Collaboration},
  journal = {Phys. Rev. X},
  volume = {13},
  issue = {4},
  pages = {041021},
  numpages = {14},
  year = {2023},
  month = {Oct},
  publisher = {American Physical Society},
  doi = {10.1103/PhysRevX.13.041021},
  url = {https://link.aps.org/doi/10.1103/PhysRevX.13.041021}
}

@article{Acernese_2023,
  title = {Frequency-Dependent Squeezed Vacuum Source for the Advanced Virgo Gravitational-Wave Detector},
  author = {Acernese, F. and Agathos, M. and Ain, A. and Albanesi, S. and All\'en\'e, C. and Allocca, A. and Amato, A. and Amra, C. and Andia, M. and Andrade, T. and Andres, N. and Andr\'es-Carcasona, M. and Andri\ifmmode \acute{c}\else \'{c}\fi{}, T. and Ansoldi, S. and Antier, S. and Apostolatos, T. and Appavuravther, E. Z. and Ar\`ene, M. and Arnaud, N. and Assiduo, M. and Melo, S. Assis de Souza and Astone, P. and Aubin, F. and Babak, S. and Badaracco, F. and Bagnasco, S. and Baird, J. and Baka, T. and Ballardin, G. and Baltus, G. and Banerjee, B. and Barneo, P. and Barone, F. and Barsuglia, M. and Barta, D. and Basti, A. and Bawaj, M. and Bazzan, M. and Beirnaert, F. and Bejger, M. and Benedetto, V. and Berbel, M. and Bernuzzi, S. and Bersanetti, D. and Bertolini, A. and Bhardwaj, U. and Bianchi, A. and Bilicki, M. and Bini, S. and Bischi, M. and Bitossi, M. and Bizouard, M.-A. and Bobba, F. and Bo\"er, M. and Bogaert, G. and Boileau, G. and Boldrini, M. and Bonavena, L. D. and Bondarescu, R. and Bondu, F. and Bonnand, R. and Boschi, V. and Boudart, V. and Bouffanais, Y. and Bozzi, A. and Bradaschia, C. and Braglia, M. and Branchesi, M. and Breschi, M. and Briant, T. and Brillet, A. and Brooks, J. and Bruno, G. and Bucci, F. and Bulashenko, O. and Bulik, T. and Bulten, H. J. and Buscicchio, R. and Buskulic, D. and Buy, C. and Cabras, G. and Cabrita, R. and Cagnoli, G. and Calloni, E. and Canepa, M. and Santoro, G. Caneva and Cannavacciuolo, M. and Capocasa, E. and Carapella, G. and Carbognani, F. and Carpinelli, M. and Carullo, G. and Diaz, J. Casanueva and Casentini, C. and Caudill, S. and Cavalieri, R. and Cella, G. and Cerd\'a-Dur\'an, P. and Cesarini, E. and Chaibi, W. and Chanial, P. and Chassande-Mottin, E. and Chaty, S. and Chessa, P. and Chiadini, F. and Chiarini, G. and Chierici, R. and Chincarini, A. and Chiofalo, M. L. and Chiummo, A. and Christensen, N. and Chua, S. and Ciani, G. and Ciecielag, P. and Cie\ifmmode \acute{s}\else \'{s}\fi{}lar, M. and Cifaldi, M. and Ciolfi, R. and Clesse, S. and Cleva, F. and Coccia, E. and Codazzo, E. and Cohadon, P.-F. and Colombo, A. and Colpi, M. and Conti, L. and Cordero-Carri\'on, I. and Corezzi, S. and Cortese, S. and Coulon, J.-P. and Coupechoux, J.-F. and Croquette, M. and Cudell, J. R. and Cuoco, E. and Cury\l{}o, M. and Dabadie, P. and Canton, T. Dal and Dall'Osso, S. and D\'alya, G. and D'Angelo, B. and Dangoisse, G. and Danilishin, S. and D'Antonio, S. and Dattilo, V. and Davier, M. and Degallaix, J. and De Laurentis, M. and Del\'eglise, S. and De Lillo, F. and Dell'Aquila, D. and Del Pozzo, W. and De Matteis, F. and Depasse, A. and De Pietri, R. and De Rosa, R. and De Rossi, C. and De Simone, R. and Di Fiore, L. and Di Giorgio, C. and Di Giovanni, F. and Di Giovanni, M. and Di Girolamo, T. and Diksha, D. and Di Lieto, A. and Di Michele, A. and Ding, J. and Di Pace, S. and Di Palma, I. and Di Renzo, F. and D'Onofrio, L. and Dooney, T. and Dorosh, O. and Drago, M. and Ducoin, J.-G. and Dupletsa, U. and Durante, O. and D'Urso, D. and Duverne, P.-A. and Eisenmann, M. and Errico, L. and Estevez, D. and Fabrizi, F. and Faedi, F. and Fafone, V. and Favaro, G. and Fays, M. and Fenyvesi, E. and Ferrante, I. and Fidecaro, F. and Figura, P. and Fiori, A. and Fiori, I. and Fittipaldi, R. and Fiumara, V. and Flaminio, R. and Font, J. A. and Frasca, S. and Frasconi, F. and Freise, A. and Freitas, O. and Fronz\'e, G. G. and Gadre, B. and Gamba, R. and Garaventa, B. and Garcia-Bellido, J. and Gargiulo, J. and Garufi, F. and Gasbarra, C. and Gemme, G. and Gennai, A. and Ghosh, Archisman and Giacoppo, L. and Giri, P. and Gissi, F. and Gkaitatzis, S. and Glotin, F. and Goncharov, B. and Gosselin, M. and Gouaty, R. and Grado, A. and Granata, M. and Granata, V. and Greco, G. and Grignani, G. and Grimaldi, A. and Guerra, D. and Guetta, D. and Guidi, G. M. and Gulminelli, F. and Guo, Y. and Gupta, P. and Gutierrez, N. and Haegel, L. and Halim, O. and Hannuksela, O. and Harder, T. and Haris, K. and Harmark, T. and Harms, J. and Haskell, B. and Heidmann, A. and Heitmann, H. and Hello, P. and Hemming, G. and Hennes, E. and Hennig, J.-S. and Hennig, M. and Hild, S. and Hofman, D. and Holland, N. A. and Hui, V. and Iandolo, G. A. and Idzkowski, B. and Iess, A. and Iorio, G. and Iosif, P. and Jacqmin, T. and Jacquet, P.-E. and Janquart, J. and Janssens, K. and Jaraba, S. and Jaranowski, P. and Jasal, P. and Juste, V. and Kalaghatgi, C. and Karathanasis, C. and Katsanevas, S. and K\'ef\'elian, F. and Koekoek, G. and Koley, S. and Kolstein, M. and Kranzhoff, S. L. and Kr\'olak, A. and Kuijer, P. and Kuroyanagi, S. and Lagabbe, P. and Laghi, D. and Lalleman, M. and Lamberts, A. and La Rana, A. and La Rosa, I. and Lartaux-Vollard, A. and Lazzaro, C. and Leaci, P. and Lema\^{\i}tre, A. and Lenti, M. and Leonova, E. and Lequime, M. and Leroy, N. and Letendre, N. and Lethuillier, M. and Leyde, K. and Linde, F. and London, L. and Longo, A. and Portilla, M. Lopez and Lorenzini, M. and Loriette, V. and Losurdo, G. and Lumaca, D. and Macquet, A. and Magazz\`u, C. and Maggiore, R. and Magnozzi, M. and Majorana, E. and Man, N. and Mangano, V. and Mantovani, M. and Mapelli, M. and Marchesoni, F. and Pina, D. Mar\'{\i}n and Marion, F. and Marquina, A. and Marsat, S. and Martelli, F. and Martinez, M. and Martinez, V. and Masserot, A. and Mastrodicasa, M. and Mastrogiovanni, S. and Meijer, Q. and Menendez-Vazquez, A. and Mereni, L. and Merzougui, M. and Miani, A. and Michel, C. and Miller, A. and Miller, B. and Milotti, E. and Minenkov, Y. and Mir, Ll. M. and Miravet-Ten\'es, M. and Mitchell, A. L. and Mondal, C. and Montani, M. and Morawski, F. and Morras, G. and Moscatello, A. and Mours, B. and Mow-Lowry, C. M. and Msihid, E. and Muciaccia, F. and Mukherjee, Suvodip and Nagar, A. and Napolano, V. and Nardecchia, I. and Narola, H. and Naticchioni, L. and Neilson, J. and Nesseris, S. and Nguyen, C. and Nieradka, G. and Nissanke, S. and Nitoglia, E. and Nocera, F. and Novak, J. and no Siles, J. F. Nu and Oertel, M. and Oganesyan, G. and Oliveri, R. and Orselli, M. and Palomba, C. and Pang, P. T. H. and Pannarale, F. and Paoletti, F. and Paoli, A. and Paolone, A. and Pappas, G. and Parisi, A. and Pascucci, D. and Pasqualetti, A. and Passaquieti, R. and Passuello, D. and Patricelli, B. and Pedurand, R. and Pegna, R. and Pegoraro, M. and Perego, A. and Pereira, A. and P\'erigois, C. and Perreca, A. and Perri\`es, S. and Perry, J. W. and Pesios, D. and Petrillo, C. and Phukon, K. S. and Piccinni, O. J. and Pichot, M. and Piendibene, M. and Piergiovanni, F. and Pierini, L. and Pierra, G. and Pierro, V. and Pillant, G. and Pillas, M. and Pilo, F. and Pinard, L. and Pinto, I. M. and Pinto, M. and Pinto, M. and Piotrzkowski, K. and Placidi, A. and Placidi, E. and Plastino, W. and Poggiani, R. and Polini, E. and Porcelli, E. and Portell, J. and Porter, E. K. and Poulton, R. and Pracchia, M. and Pradier, T. and Principe, M. and Prodi, G. A. and Prosposito, P. and Puecher, A. and Punturo, M. and Puosi, F. and Puppo, P. and Raaijmakers, G. and Radulesco, N. and Rapagnani, P. and Razzano, M. and Regimbau, T. and Rei, L. and Rettegno, P. and Revenu, B. and Reza, A. and Rezaei, A. S. and Ricci, F. and Rinaldi, S. and Robinet, F. and Rocchi, A. and Rolland, L. and Romanelli, M. and Romano, R. and Romero, A. and Ronchini, S. and Rosa, L. and Rosi\ifmmode \acute{n}\else \'{n}\fi{}ska, D. and Roy, S. and Rozza, D. and Ruggi, P. and Morales, E. Ruiz and Saffarieh, P. and Salafia, O. S. and Salconi, L. and Salemi, F. and Sall\'e, M. and Samajdar, A. and Sanchis-Gual, N. and Sanuy, A. and Sasli, A. and Sassi, P. and Sassolas, B. and Sayah, S. and Schmidt, S. and Seglar-Arroyo, M. and Sentenac, D. and Sequino, V. and Servignat, G. and Setyawati, Y. and Shcheblanov, N. S. and Sieniawska, M. and Silenzi, L. and Singh, N. and Singha, A. and Sipala, V. and Soldateschi, J. and Sordini, V. and Sorrentino, F. and Sorrentino, N. and Soulard, R. and Spagnuolo, V. and Spera, M. and Spinicelli, P. and Stachie, C. and Steer, D. A. and Steinlechner, J. and Steinlechner, S. and Stergioulas, N. and Stratta, G. and Suchenek, M. and Sur, A. and Suresh, J. and Swinkels, B. L. and Syx, A. and Szewczyk, P. and Tacca, M. and Tamanini, N. and Tanasijczuk, A. J. and Mart\'{\i}n, E. N. Tapia San and Taranto, C. and Tonelli, M. and Torres-Forn\'e, A. and e Melo, I. Tosta and Tournefier, E. and Trapananti, A. and Travasso, F. and Trenado, J. and Tringali, M. C. and Troiano, L. and Trovato, A. and Trozzo, L. and Tsang, K. W. and Turbang, K. and Turconi, M. and Turski, C. and Ubach, H. and Utina, A. and Valentini, M. and Vallero, S. and van Bakel, N. and van Beuzekom, M. and van Dael, M. and van den Brand, J. F. J. and Van Den Broeck, C. and van der Sluys, M. and Van de Walle, A. and van Dongen, J. and van Haevermaet, H. and van Heijningen, J. V. and van Ranst, Z. and van Remortel, N. and Vardaro, M. and Vas\'uth, M. and Vedovato, G. and Verdier, P. and Verkindt, D. and Verma, P. and Vetrano, F. and Vicer\'e, A. and Vinet, J.-Y. and Viret, S. and Virtuoso, A. and Vocca, H. and Walet, R. C. and Was, M. and Yadav, N. and Zadro\ifmmode \dot{z}\else \.{z}\fi{}ny, A. and Zelenova, T. and Zendri, J.-P. and Zhao, Y. and Zerrad, M. and Vahlbruch, H. and Mehmet, M. and L\"uck, H. and Danzmann, K.},
  collaboration = {Virgo Collaboration},
  journal = {Phys. Rev. Lett.},
  volume = {131},
  issue = {4},
  pages = {041403},
  numpages = {11},
  year = {2023},
  month = {Jul},
  publisher = {American Physical Society},
  doi = {10.1103/PhysRevLett.131.041403},
  url = {https://link.aps.org/doi/10.1103/PhysRevLett.131.041403}
}

@article{McCuller_2021,
  title = {LIGO's quantum response to squeezed states},
  author = {McCuller, L. and Dwyer, S. E. and Green, A. C. and Yu, Haocun and Kuns, K. and Barsotti, L. and Blair, C. D. and Brown, D. D. and Effler, A. and Evans, M. and Fernandez-Galiana, A. and Fritschel, P. and Frolov, V. V. and Kijbunchoo, N. and Mansell, G. L. and Matichard, F. and Mavalvala, N. and McClelland, D. E. and McRae, T. and Mullavey, A. and Sigg, D. and Slagmolen, B. J. J. and Tse, M. and Vo, T. and Ward, R. L. and Whittle, C. and Abbott, R. and Adams, C. and Adhikari, R. X. and Ananyeva, A. and Appert, S. and Arai, K. and Areeda, J. S. and Asali, Y. and Aston, S. M. and Austin, C. and Baer, A. M. and Ball, M. and Ballmer, S. W. and Banagiri, S. and Barker, D. and Bartlett, J. and Berger, B. K. and Betzwieser, J. and Bhattacharjee, D. and Billingsley, G. and Biscans, S. and Blair, R. M. and Bode, N. and Booker, P. and Bork, R. and Bramley, A. and Brooks, A. F. and Buikema, A. and Cahillane, C. and Cannon, K. C. and Chen, X. and Ciobanu, A. A. and Clara, F. and Compton, C. M. and Cooper, S. J. and Corley, K. R. and Countryman, S. T. and Covas, P. B. and Coyne, D. C. and Datrier, L. E. H. and Davis, D. and Di Fronzo, C. and Dooley, K. L. and Driggers, J. C. and Etzel, T. and Evans, T. M. and Feicht, J. and Fulda, P. and Fyffe, M. and Giaime, J. A. and Giardina, K. D. and Godwin, P. and Goetz, E. and Gras, S. and Gray, C. and Gray, R. and Gustafson, E. K. and Gustafson, R. and Hanks, J. and Hanson, J. and Hardwick, T. and Hasskew, R. K. and Heintze, M. C. and Helmling-Cornell, A. F. and Holland, N. A. and Jones, J. D. and Kandhasamy, S. and Karki, S. and Kasprzack, M. and Kawabe, K. and King, P. J. and Kissel, J. S. and Kumar, Rahul and Landry, M. and Lane, B. B. and Lantz, B. and Laxen, M. and Lecoeuche, Y. K. and Leviton, J. and Liu, J. and Lormand, M. and Lundgren, A. P. and Macas, R. and MacInnis, M. and Macleod, D. M. and M\'arka, S. and M\'arka, Z. and Martynov, D. V. and Mason, K. and Massinger, T. J. and McCarthy, R. and McCormick, S. and McIver, J. and Mendell, G. and Merfeld, K. and Merilh, E. L. and Meylahn, F. and Mistry, T. and Mittleman, R. and Moreno, G. and Mow-Lowry, C. M. and Mozzon, S. and Nelson, T. J. N. and Nguyen, P. and Nuttall, L. K. and Oberling, J. and Oram, Richard J. and Osthelder, C. and Ottaway, D. J. and Overmier, H. and Palamos, J. R. and Parker, W. and Payne, E. and Pele, A. and Penhorwood, R. and Perez, C. J. and Pirello, M. and Radkins, H. and Ramirez, K. E. and Richardson, J. W. and Riles, K. and Robertson, N. A. and Rollins, J. G. and Romel, C. L. and Romie, J. H. and Ross, M. P. and Ryan, K. and Sadecki, T. and Sanchez, E. J. and Sanchez, L. E. and Saravanan, T. R. and Savage, R. L. and Schaetzl, D. and Schnabel, R. and Schofield, R. M. S. and Schwartz, E. and Sellers, D. and Shaffer, T. and Smith, J. R. and Soni, S. and Sorazu, B. and Spencer, A. P. and Strain, K. A. and Sun, L. and Szczepa\ifmmode \acute{n}\else \'{n}\fi{}czyk, M. J. and Thomas, M. and Thomas, P. and Thorne, K. A. and Toland, K. and Torrie, C. I. and Traylor, G. and Urban, A. L. and Vajente, G. and Valdes, G. and Vander-Hyde, D. C. and Veitch, P. J. and Venkateswara, K. and Venugopalan, G. and Viets, A. D. and Vorvick, C. and Wade, M. and Warner, J. and Weaver, B. and Weiss, R. and Willke, B. and Wipf, C. C. and Xiao, L. and Yamamoto, H. and Yu, Hang and Zhang, L. and Zucker, M. E. and Zweizig, J.},
  journal = {Phys. Rev. D},
  volume = {104},
  issue = {6},
  pages = {062006},
  numpages = {29},
  year = {2021},
  month = {Sep},
  publisher = {American Physical Society},
  doi = {10.1103/PhysRevD.104.062006},
  url = {https://link.aps.org/doi/10.1103/PhysRevD.104.062006}
}

@article{Toyra_2017,
  title = {Multi-spatial-mode effects in squeezed-light-enhanced interferometric gravitational wave detectors},
  author = {T\"oyr\"a, Daniel and Brown, Daniel D. and Davis, McKenna and Song, Shicong and Wormald, Alex and Harms, Jan and Miao, Haixing and Freise, Andreas},
  journal = {Phys. Rev. D},
  volume = {96},
  issue = {2},
  pages = {022006},
  numpages = {11},
  year = {2017},
  month = {Jul},
  publisher = {American Physical Society},
  doi = {10.1103/PhysRevD.96.022006},
  url = {https://link.aps.org/doi/10.1103/PhysRevD.96.022006}
}

@article{Morrison:94,
author = {Euan Morrison and Brian J. Meers and David I. Robertson and Henry Ward},
journal = {Appl. Opt.},
number = {22},
pages = {5037--5040},
publisher = {Optica Publishing Group},
title = {Experimental demonstration of an automatic alignment system for optical interferometers},
volume = {33},
month = {Aug},
year = {1994},
url = {https://opg.optica.org/ao/abstract.cfm?URI=ao-33-22-5037},
doi = {10.1364/AO.33.005037},
}

@article{Fabian_2019,
  title = {Sensing optical cavity mismatch with a mode-converter and quadrant photodiode},
  author = {Maga\~na-Sandoval, Fabian and Vo, Thomas and Vander-Hyde, Daniel and Sanders, J. R. and Ballmer, Stefan W.},
  journal = {Phys. Rev. D},
  volume = {100},
  issue = {10},
  pages = {102001},
  numpages = {12},
  year = {2019},
  month = {Nov},
  publisher = {American Physical Society},
  doi = {10.1103/PhysRevD.100.102001},
  url = {https://link.aps.org/doi/10.1103/PhysRevD.100.102001}
}

@article{Fulda_17,
author = {P. Fulda and D. Voss and C. Mueller and L. F. Ortega and G. Ciani and G. Mueller and D. B. Tanner},
journal = {Appl. Opt.},
number = {13},
pages = {3879--3888},
publisher = {Optica Publishing Group},
title = {Alignment sensing for optical cavities using radio-frequency jitter modulation},
volume = {56},
month = {May},
year = {2017},
url = {https://opg.optica.org/ao/abstract.cfm?URI=ao-56-13-3879},
doi = {10.1364/AO.56.003879},
}

@article{Tao_25,
author = {Liu Tao and Mauricio Diaz-Ortiz and Paul Fulda},
journal = {Appl. Opt.},
number = {6},
pages = {1556--1564},
publisher = {Optica Publishing Group},
title = {High-efficiency electro-optic lens for radio-frequency beam wavefront modulation for mode mismatch sensing},
volume = {64},
month = {Feb},
year = {2025},
url = {https://opg.optica.org/ao/abstract.cfm?URI=ao-64-6-1556},
doi = {10.1364/AO.546199},
}

@article{Cao_20,
author = {Huy Tuong Cao and Daniel D. Brown and Peter J. Veitch and David J. Ottaway},
journal = {Opt. Express},
number = {10},
pages = {14405--14413},
publisher = {Optica Publishing Group},
title = {Optical lock-in camera for gravitational wave detectors},
volume = {28},
month = {May},
year = {2020},
url = {https://opg.optica.org/oe/abstract.cfm?URI=oe-28-10-14405},
doi = {10.1364/OE.384754},
}

@article{Agatsuma_19,
author = {Kazuhiro Agatsuma and Laura van der Schaaf and Martin van Beuzekom and David Rabeling and Jo van den Brand},
journal = {Opt. Express},
number = {13},
pages = {18533--18548},
publisher = {Optica Publishing Group},
title = {High-performance phase camera as a frequency selective laser wavefront sensor for gravitational wave detectors},
volume = {27},
month = {Jun},
year = {2019},
url = {https://opg.optica.org/oe/abstract.cfm?URI=oe-27-13-18533},
doi = {10.1364/OE.27.018533},
}

@article{Schiworski_21,
author = {Mitchell G. Schiworski and Daniel D. Brown and David J. Ottaway},
journal = {J. Opt. Soc. Am. A},
number = {11},
pages = {1603--1611},
publisher = {Optica Publishing Group},
title = {Modal decomposition of complex optical fields using convolutional neural networks},
volume = {38},
month = {Nov},
year = {2021},
url = {https://opg.optica.org/josaa/abstract.cfm?URI=josaa-38-11-1603},
doi = {10.1364/JOSAA.428214},
}

@misc{tao2025FLIR,
      title={Error signals for overcoming the laser power limits of gravitational-wave detectors}, 
      author={Liu Tao and Pooyan Goodarzi and Jonathan W. Richardson},
      year={2025},
      eprint={2509.06840},
      archivePrefix={arXiv},
      primaryClass={astro-ph.IM},
      url={https://arxiv.org/abs/2509.06840}, 
}

@article{Tao_21_loss,
author = {Liu Tao and Jessica Kelley-Derzon and Anna C. Green and Paul Fulda},
journal = {Opt. Lett.},
number = {11},
pages = {2694--2697},
publisher = {Optica Publishing Group},
title = {Power coupling losses for misaligned and mode-mismatched higher-order Hermite--Gauss modes},
volume = {46},
month = {Jun},
year = {2021},
url = {https://opg.optica.org/ol/abstract.cfm?URI=ol-46-11-2694},
doi = {10.1364/OL.426999},
}

@article{An_19,
author = {Yi An and Liangjin Huang and Jun Li and Jinyong Leng and Lijia Yang and Pu Zhou},
journal = {Opt. Express},
number = {7},
pages = {10127--10137},
publisher = {Optica Publishing Group},
title = {Learning to decompose the modes in few-mode fibers with deep convolutional neural network},
volume = {27},
month = {Apr},
year = {2019},
url = {https://opg.optica.org/oe/abstract.cfm?URI=oe-27-7-10127},
doi = {10.1364/OE.27.010127},
}

@article{An_20,
author = {Yi An and Tianyue Hou and Jun Li and Liangjin Huang and Jinyong Leng and Lijia Yang and Pu Zhou},
journal = {Appl. Opt.},
number = {7},
pages = {1954--1959},
publisher = {Optica Publishing Group},
title = {Fast modal analysis for Hermite--Gaussian beams via deep learning},
volume = {59},
month = {Mar},
year = {2020},
url = {https://opg.optica.org/ao/abstract.cfm?URI=ao-59-7-1954},
doi = {10.1364/AO.377189},
}

@article{Cutolo_95,
author = {Antonello Cutolo and Tommaso Isernia and IIdegonda Izzo and Rocco Pierri and Luigi Zeni},
journal = {Appl. Opt.},
number = {34},
pages = {7974--7978},
publisher = {Optica Publishing Group},
title = {Transverse mode analysis of a laser beam by near- and far-field intensity measurements},
volume = {34},
month = {Dec},
year = {1995},
url = {https://opg.optica.org/ao/abstract.cfm?URI=ao-34-34-7974},
doi = {10.1364/AO.34.007974},
}

@article{Hofer_19,
author = {L. R. Hofer and L. W. Jones and J. L. Goedert and R. V. Dragone},
journal = {J. Opt. Soc. Am. A},
number = {6},
pages = {936--943},
publisher = {Optica Publishing Group},
title = {Hermite--Gaussian mode detection via convolution neural networks},
volume = {36},
month = {Jun},
year = {2019},
url = {https://opg.optica.org/josaa/abstract.cfm?URI=josaa-36-6-936},
doi = {10.1364/JOSAA.36.000936},
}

@article{Dong_2023,
   title={Phase Retrieval: From Computational Imaging to Machine Learning: A tutorial},
   volume={40},
   ISSN={1558-0792},
   url={http://dx.doi.org/10.1109/MSP.2022.3219240},
   DOI={10.1109/msp.2022.3219240},
   number={1},
   journal={IEEE Signal Processing Magazine},
   publisher={Institute of Electrical and Electronics Engineers (IEEE)},
   author={Dong, Jonathan and Valzania, Lorenzo and Maillard, Antoine and Pham, Thanh-an and Gigan, Sylvain and Unser, Michael},
   year={2023},
   month=jan, pages={45–57} }

@article{Bond2017,
  title = {Interferometer techniques for gravitational-wave detection},
  author = {Bond, Charlotte and Brown, Daniel and Freise, Andreas and Strain, Kenneth A},
  journal = {Living Reviews in Relativity},
  volume = {19},
  issue = {1},
  year = {2017},
  month = {Feb},
  publisher = {American Physical Society},
  doi = {10.1007/s41114-016-0002-8},
  url = {https://doi.org/10.1007/s41114-016-0002-8}
}

@misc{VGG16,
      title={Very Deep Convolutional Networks for Large-Scale Image Recognition}, 
      author={Karen Simonyan and Andrew Zisserman},
      year={2015},
      eprint={1409.1556},
      archivePrefix={arXiv},
      primaryClass={cs.CV},
      url={https://arxiv.org/abs/1409.1556}, 
}

@INPROCEEDINGS{ImageNet,
  author={Deng, Jia and Dong, Wei and Socher, Richard and Li, Li-Jia and Kai Li and Li Fei-Fei},
  booktitle={2009 IEEE Conference on Computer Vision and Pattern Recognition}, 
  title={ImageNet: A large-scale hierarchical image database}, 
  year={2009},
  volume={},
  number={},
  pages={248-255},
  doi={10.1109/CVPR.2009.5206848}}

@article{Mannam:22,
author = {Varun Mannam and Yide Zhang and Yinhao Zhu and Evan Nichols and Qingfei Wang and Vignesh Sundaresan and Siyuan Zhang and Cody Smith and Paul W. Bohn and Scott S. Howard},
journal = {Optica},
number = {4},
pages = {335--345},
publisher = {Optica Publishing Group},
title = {Real-time image denoising of mixed Poisson--Gaussian noise in fluorescence microscopy images using ImageJ},
volume = {9},
month = {Apr},
year = {2022},
url = {https://opg.optica.org/optica/abstract.cfm?URI=optica-9-4-335},
doi = {10.1364/OPTICA.448287},
}

@article{GWD_noise_ML,
  title = {Machine-learning nonstationary noise out of gravitational-wave detectors},
  author = {Vajente, G. and Huang, Y. and Isi, M. and Driggers, J. C. and Kissel, J. S. and Szczepa\ifmmode \acute{n}\else \'{n}\fi{}czyk, M. J. and Vitale, S.},
  journal = {Phys. Rev. D},
  volume = {101},
  issue = {4},
  pages = {042003},
  numpages = {12},
  year = {2020},
  month = {Feb},
  publisher = {American Physical Society},
  doi = {10.1103/PhysRevD.101.042003},
  url = {https://link.aps.org/doi/10.1103/PhysRevD.101.042003}
}

@misc{soni2025gwyolo,
      title={GW-YOLO: Multi-transient segmentation in LIGO using computer vision}, 
      author={Siddharth Soni and Nikhil Mukund and Erik Katsavounidis},
      year={2025},
      eprint={2508.17399},
      archivePrefix={arXiv},
      primaryClass={astro-ph.IM},
      url={https://arxiv.org/abs/2508.17399}, 
}

@article{Goodwin-Jones:24,
author = {Aaron W. Goodwin-Jones and Ricardo Cabrita and Mikhail Korobko and Martin Van Beuzekom and Daniel D. Brown and Viviana Fafone and Joris Van Heijningen and Alessio Rocchi and Mitchell G. Schiworski and Matteo Tacca},
journal = {Optica},
number = {2},
pages = {273--290},
publisher = {Optica Publishing Group},
title = {Transverse mode control in quantum enhanced interferometers: a review and recommendations for a new generation},
volume = {11},
month = {Feb},
year = {2024},
url = {https://opg.optica.org/optica/abstract.cfm?URI=optica-11-2-273},
doi = {10.1364/OPTICA.511924},
}

@article{Qin_2025,
doi = {10.1088/1361-6382/ada864},
url = {https://doi.org/10.1088/1361-6382/ada864},
year = {2025},
month = {jan},
publisher = {IOP Publishing},
volume = {42},
number = {4},
pages = {045003},
author = {Qin, Jiayi and Kinder, Katherine and Jadhav, Shreejit and Chugh, Praneel and Slagmolen, Bram J J},
title = {Automated alignment of an optical cavity using machine learning},
journal = {Classical and Quantum Gravity}
}

@article{Mukund2023,
  title = {Neural sensing and control in a kilometer-scale gravitational-wave observatory},
  author = {Mukund, N. and Lough, J. and Bisht, A. and Wittel, H. and Nadji, S. and Affeldt, C. and Bergamin, F. and Brinkmann, M. and Kringel, V. and L\"uck, H. and Weinert, M. and Danzmann, K.},
  journal = {Phys. Rev. Appl.},
  volume = {20},
  issue = {6},
  pages = {064041},
  numpages = {11},
  year = {2023},
  month = {Dec},
  publisher = {American Physical Society},
  doi = {10.1103/PhysRevApplied.20.064041},
  url = {https://link.aps.org/doi/10.1103/PhysRevApplied.20.064041}
}

@article{Wang:21,
author = {Meng Wang and Yuanyuan Ma and Quan Sheng and Xi He and Junjie Liu and Wei Shi and Jianquan Yao and Takashige Omatsu},
journal = {Opt. Express},
number = {17},
pages = {27783--27790},
publisher = {Optica Publishing Group},
title = {Laguerre-Gaussian beam generation via enhanced intracavity spherical aberration},
volume = {29},
month = {Aug},
year = {2021},
url = {https://opg.optica.org/oe/abstract.cfm?URI=oe-29-17-27783},
doi = {10.1364/OE.436110},
}

@article{PhysRevD.111.062002,
  title = {Advanced LIGO detector performance in the fourth observing run},
  author = {Capote, E. and Jia, W. and Aritomi, N. and Nakano, M. and Xu, V. and Abbott, R. and Abouelfettouh, I. and Adhikari, R. X. and Ananyeva, A. and Appert, S. and Apple, S. K. and Arai, K. and Aston, S. M. and Ball, M. and Ballmer, S. W. and Barker, D. and Barsotti, L. and Berger, B. K. and Betzwieser, J. and Bhattacharjee, D. and Billingsley, G. and Biscans, S. and Blair, C. D. and Bode, N. and Bonilla, E. and Bossilkov, V. and Branch, A. and Brooks, A. F. and Brown, D. D. and Bryant, J. and Cahillane, C. and Cao, H. and Clara, F. and Collins, J. and Compton, C. M. and Cottingham, R. and Coyne, D. C. and Crouch, R. and Csizmazia, J. and Cumming, A. and Dartez, L. P. and Davis, D. and Demos, N. and Dohmen, E. and Driggers, J. C. and Dwyer, S. E. and Effler, A. and Ejlli, A. and Etzel, T. and Evans, M. and Feicht, J. and Frey, R. and Frischhertz, W. and Fritschel, P. and Frolov, V. V. and Fuentes-Garcia, M. and Fulda, P. and Fyffe, M. and Ganapathy, D. and Gateley, B. and Gayer, T. and Giaime, J. A. and Giardina, K. D. and Glanzer, J. and Goetz, E. and Goetz, R. and Goodwin-Jones, A. W. and Gras, S. and Gray, C. and Griffith, D. and Grote, H. and Guidry, T. and Gurs, J. and Hall, E. D. and Hanks, J. and Hanson, J. and Heintze, M. C. and Helmling-Cornell, A. F. and Holland, N. A. and Hoyland, D. and Huang, H. Y. and Inoue, Y. and James, A. L. and Jamies, A. and Jennings, A. and Jones, D. H. and Kabagoz, H. B. and Karat, S. and Karki, S. and Kasprzack, M. and Kawabe, K. and Kijbunchoo, N. and King, P. J. and Kissel, J. S. and Komori, K. and Kontos, A. and Kumar, Rahul and Kuns, K. and Landry, M. and Lantz, B. and Laxen, M. and Lee, K. and Lesovsky, M. and Villarreal, F. Llamas and Lormand, M. and Loughlin, H. A. and Macas, R. and MacInnis, M. and Makarem, C. N. and Mannix, B. and Mansell, G. L. and Martin, R. M. and Mason, K. and Matichard, F. and Mavalvala, N. and Maxwell, N. and McCarrol, G. and McCarthy, R. and McClelland, D. E. and McCormick, S. and McRae, T. and Mera, F. and Merilh, E. L. and Meylahn, F. and Mittleman, R. and Moraru, D. and Moreno, G. and Mullavey, A. and Nelson, T. J. N. and Neunzert, A. and Notte, J. and Oberling, J. and O'Hanlon, T. and Osthelder, C. and Ottaway, D. J. and Overmier, H. and Parker, W. and Patane, O. and Pele, A. and Pham, H. and Pirello, M. and Pullin, J. and Quetschke, V. and Ramirez, K. E. and Ransom, K. and Reyes, J. and Richardson, J. W. and Robinson, M. and Rollins, J. G. and Romel, C. L. and Romie, J. H. and Ross, M. P. and Ryan, K. and Sadecki, T. and Sanchez, A. and Sanchez, E. J. and Sanchez, L. E. and Savage, R. L. and Schaetzl, D. and Schiworski, M. G. and Schnabel, R. and Schofield, R. M. S. and Schwartz, E. and Sellers, D. and Shaffer, T. and Short, R. W. and Sigg, D. and Slagmolen, B. J. J. and Soike, C. and Soni, S. and Srivastava, V. and Sun, L. and Tanner, D. B. and Thomas, M. and Thomas, P. and Thorne, K. A. and Todd, M. R. and Torrie, C. I. and Traylor, G. and Ubhi, A. S. and Vajente, G. and Vanosky, J. and Vecchio, A. and Veitch, P. J. and Vibhute, A. M. and von Reis, E. R. G. and Warner, J. and Weaver, B. and Weiss, R. and Whittle, C. and Willke, B. and Wipf, C. C. and Wright, J. L. and Yamamoto, H. and Zhang, L. and Zucker, M. E.},
  journal = {Phys. Rev. D},
  volume = {111},
  issue = {6},
  pages = {062002},
  numpages = {30},
  year = {2025},
  month = {Mar},
  publisher = {American Physical Society},
  doi = {10.1103/PhysRevD.111.062002},
  url = {https://link.aps.org/doi/10.1103/PhysRevD.111.062002}
}

@Article{photonics12080801,
AUTHOR = {Chen, Tai and Jiang, Chengcai and Tao, Jia and Ma, Long and Cao, Longzhou},
TITLE = {Generation of Higher-Order Hermite–Gaussian Modes Based on Physical Model and Deep Learning},
JOURNAL = {Photonics},
VOLUME = {12},
YEAR = {2025},
NUMBER = {8},
ARTICLE-NUMBER = {801},
URL = {https://www.mdpi.com/2304-6732/12/8/801},
ISSN = {2304-6732},
DOI = {10.3390/photonics12080801}
}

@misc{diab2025quantitativeperformanceanalysisdifferent,
      title={A quantitative performance analysis of two different interferometric alignment sensing schemes for gravitational wave detectors}, 
      author={Raed Diab and Alvaro Herrera and Chance Jackson and Paul Fulda},
      year={2025},
      eprint={2510.09774},
      archivePrefix={arXiv},
      primaryClass={astro-ph.IM},
      url={https://arxiv.org/abs/2510.09774}, 
}

@misc{kingma2017adammethodstochasticoptimization,
      title={Adam: A Method for Stochastic Optimization}, 
      author={Diederik P. Kingma and Jimmy Ba},
      year={2017},
      eprint={1412.6980},
      archivePrefix={arXiv},
      primaryClass={cs.LG},
      url={https://arxiv.org/abs/1412.6980}, 
}

\end{document}